\documentstyle[preprint,aps,prb,epsf]{revtex}

\newcommand{\mb}[1]{ { \mbox{\boldmath{$#1$}}}  }
\newcommand{\mbs}[1]{ {\scriptsize \mbox{\boldmath{$#1$}}}  }
\newcommand{\sub}[2]{ \mbox{$#1$}_{\mbox{\scriptsize\boldmath{$#2$}}}}
\begin{document}
\draft
\title{ Van Hove Singularity and Superconductivity 
in Disordered Hubbard
Model 
}

\author{Grzegorz Litak}

\address{
Department of Mechanics, Technical University of
Lublin \\ Nadbystrzycka 36, PL--20-618  Lublin, Poland.}

\date{\today}
\maketitle

\begin{abstract}
We apply the Coherent Potential Approximation (CPA) to 
a simple extended Hubbard model with a nearest  and next nearest
neighbour hopping for disordered superconductors with $s$--, $d$--
and $p$--wave
pairing. 
We show how the  Van Hove 
singularities in the electron density of states enhance the transition
temperature
$T_c$
for exotic superconductors in a clean and weakly disordered
system. The Anderson theorem and pair--breaking effects in presence
of Van Hove singularity caused by
non--magnetic
disorder are also discussed.
\end{abstract}
\pacs{Pacs. 74.62.Dh, 74.20-z}


\section{Introduction}

Magnetic and non--magnetic
impurities in superconductors always attracted interest and  their
treatment played
the essential
role in
theories of
superconductivity.

Since the Anderson \cite{And59} and Abrikosov, Gorkov
 \cite{Abr59} works
 the
influence
of magnetic and non--magnetic disorder on superconductors have been
treated in many ways \cite{Mak69,Sun95,Mak00,Har98,Har00,Bel94,Gho01}.
Their
arguments, originally
applied to
classic BCS
superconductors, were reexamined for
novel, exotic superconductors \cite{Ann96,Ann00} 
 with the anisotropic order parameters
extended $s$--wave, $d$--wave character
\cite{Sun95,Gor83,Pok96,Ner94,Ner95,Ner97,Feh96,Wen95,Lit98,Mar99,Rad93,Den98,Ope97,Bor94,Ope98,Buh00,Lok01}
and
also  
recently discovered \cite{Mae94,Bas96,Agt97,Mae01} 
$p$--wave
ruthenates \cite{Mak00,Miy98,Puc00,Lit00b,Lit01}

 Examining  the influence of various kinds of
magnetic and  non--magnetic disorder caused by
 structural, substitutional, irradiation defects  etc. on
properties of conventional  and unconventional superconductors
 has shown that  their responses  to
disorder are quite different. 
In contrast to conventional materials
where only
magnetic impurities affect the superconducting properties, for
unconventional superconductors the effect of
both magnetic and non-magnetic disorder is usually strong
\cite{Wes94,Pen94,Uch95,Ulm95,Xia90,Ber96,Tal97,Kuo97,Wil98,Gia94,Ele96,Cie98,Kar97,Kar00,Mao99,Mac98}. 

Thus, it is not surprising, that the response  to disorder become the
fundamental criterion of unconventionality of the  physical  mechanism
leading
to
superconductivity. Moreover, effects of disorder are of interest because the  high
temperature superconductors have to be
doped
(La$_{2-x}$
Sr$_x$CuO$_4$
with Strontium $x=0.15$,
YBa$_2$Cu$_3$O$_{7-x}$ with Oxygen $x=0.1$) to show the
optimal critical temperature and doping is always accompanied by disorder.

On the other hand in these quasi two dimensional
layered   
materials 
the Fermi energy
has been found in the vicinity of Van Hove singularities of
the electron
density of states
\cite{Mar97,Fri89,Dzi87,Lab87,Tsu90,Bok97,Bou98,Bok00,Szo98,And94,Nov96,Pic97,New95,Que99,Que99a,Lu96}. 
This has lead to formulation of 
Van Hove scenario for high temperature superconductors
which 
says that optimal critical temperature is reached when chemical potential  
passes through the Van Hove singularity in the
density of states \cite{Mar97,Fri89,Bou98,Lit98}.
Doping with charge carriers does not only
change the density  
in the system but also smear the density of states
eliminating its singularities
and,  specially 
for anisotropic superconductors, introduce the electron pair--breaking
phenomenon \cite{Gor83,Sun95,Har98,Mar99}.  

Thus,  one has to investigate  
the effect of rise and fall of the critical temperature $T_c$ near the Van Hove
singularities very carefully, simultaneously, taking into account   
the effects of disorder.
Clearly disorder reduces the critical temperature $T_c$
by elimination of singularities in the density of states
and, at the same time, by  braking pairs. 
The present paper is the extension of the previous ones  \cite{Lit98,Mar99,Lit00b}
where the electron hopping was introduced only between nearest neighbour lattice sites.
Here following  Refs. \onlinecite{Mic90,Mic97} we introduce the hopping to
next
nearest sites $t'$
(i.e. the ratio of  hopping parameters $t$, $t'$ for YBaCuO was suggested to be 
$t'/t=0.45$)
and investigate the effect of disorder on superconducting properties.

In the normal states the  nonzero hopping to the next neighbour sites 
$t'$ introduces the
distortion of the Fermi surface (Fig. 1a) which  results in  considerable changes in the
electron density of states
(Fig 1b). Note, in case  $t' \ne 0$ the central Van Hove singularity produces the
stronger enhancement of
the density of states ($N(E)$ for $E \approx E_{v1}$)
than for $t'=0$ (Fig. 1b).
Moreover, the other Van Hove singularity associated with  the bottom edge of the band
$E_{v2}$ creates the second peak in the density of states
$N(E)$.
Thus, depending on the band filling, 
the additional electron hopping $t'$ should have the effect on critical temperature
$T_c$, rising it to higher value. On the other hand, the 
distorted Fermi surface with the stronger dependence on $\mb k$  (Fig. 1a)
seems to be less stable in the presence disorder as  $\mb k$ is not a good
quantum number in disordered system.

The present  paper is organized as follows. 
 In Section II  the discussion starts  with  the 
extended attractive Hubbard model defined on the square lattice which has the
extended $s$-- $d$-- and $p$--
wave order parameter
solutions. Than  we  shortly overview and classify various
types of 
Van Hove singularities in electron density of states of 
one band model and their influence
on the superconducting critical temperature $T_c$.
In Section  III we investigate disordered superconducting systems and
apply the
Coherent Potential Approximation (CPA)  for attractive Hubbard model.
The Anderson theorem for non-magnetic impurity effect on isotropic
$s$--wave solutions
and the pair--breaking effect in case of anisotropic pairing is also
discussed.
Section IV is devoted to the examination of the disorder effect on 
the superconducting critical
temperature $T_c$. Finally,
Section V contains conclusions and remarks.

\section{The role of Van Hove singularities in the clean system} 
\subsection{Bogoliubov--de Gennes Equation}

We start with the single band Hubbard model with an attractive
extended
interaction which is described by the following Hamiltonian
\cite{Mic90,Mic97,Mic88}:

\begin{equation}
\label{eq1}
H=\sum_{ij \sigma} t_{ij}c^{\dagger}_{i \sigma} c_{j \sigma} +
\frac{1}{2} \sum_{ij} U_{ij} n_i n_j - \sum_{i}(\mu-\varepsilon_i)n_i.
\end{equation}
In the above $n_i=n_{i \uparrow}+n_{i \downarrow}$ is the charge on  site
labeled $i$, $\mu$ is the
chemical
potential. Disorder is introduced into the problem by allowing the local
site
energy $\varepsilon_i$ to vary randomly from site to site, $c^{\dagger}_{i
\sigma}$ and $c_{i \sigma}$ are the Fermion
creation and annihilation operators for an electron
 on site $i$ with spin $\sigma$,
$t_{ij}$ is the amplitude for
hopping from site $j$ to site $i$ (with
$t_{ii}=0$) and finally  $U_{ij}$ is the attractive interaction ($U_{ij} <
0$)
which causes superconductivity and can be
either  local ($i=j$) or non-local ($i\ne j$).
Starting from Eq. (\ref{eq1}) we apply the Hartree--Fock--Gorkov
\cite{Tin75}
approximation, which results in the Bogoliubov--de Gennes
equation for a singlet ($s$-- or $d$-- wave) superconductor:

\begin{equation}
\label{eq2}
\sum_l \left( \begin{array}{c} (\varepsilon_i - \mu)
\delta_{il}-t_{il}~~~~-\Delta_{il} \\
 -\Delta_{il}^*~~~~ (-\varepsilon_i + \mu) \delta_{il}
+t_{il} \end{array} \right)
 \left( \begin{array}{c} u_{l \uparrow}
\\
v_{l \downarrow} \end{array} \right) =
E \left( \begin{array}{c} u_{i \uparrow}
\\
v_{i \downarrow  } \end{array} \right),
\end{equation}  
where $u_{l \uparrow}$ and $v_{l \downarrow}$ are electron and hole wave
functions with up anti parallel spins $\uparrow$ and $\downarrow$
respectively. 
The usual singlet one particle Green function, in the Nambu
space $\mb G (i,j;  \imath \omega_n)$, 
at the Matsubara frequency $\omega_n= \frac{\pi}{\beta} (2n+1)$  
satisfies:

\begin{equation}
\label{eq3}
\sum_l \left( \begin{array}{c} (\imath \omega_n -\varepsilon_i + \mu) 
\delta_{il}+t_{il}~~~~~\Delta_{il} \\ 
 \Delta_{il}^*~~~~~ (\imath \omega_n +\varepsilon_i - \mu) \delta_{il}
-t_{il} \end{array} \right)
 \left( \begin{array}{c} G_{11}(l,j;\imath \omega_n)~~
G_{12}(l,j;\imath \omega_n) \\
G_{21}(l,j;\imath \omega_n)~~ G_{22}(l,j;\imath \omega_n) \end{array}
\right)
= 
\delta_{ij} \mb 1
\end{equation}
where  the pairing potentials 
$\Delta_{ij}$ will be taken to 
be non zero only when the sites $i$ and $j$ coincide ($i=j$),
for
on site interaction $U_{ii}$, 
or are nearest neighbours, for off diagonal interaction $U_{ij}$.
On the other hand in  case of triplet, $p$--wave, paring instead the Eq.
\ref{eq2} we
have:
\begin{equation}
\label{eq4}
\sum_l \left( \begin{array}{c} (\varepsilon_i - \mu)
\delta_{il}-t_{il}~~~~~0~~~~~~~~~~~-\Delta_{il}^{\uparrow
\uparrow}~~~~~~~~~~~-\Delta_{il}^{\uparrow \downarrow} \\
0~~~~~(\varepsilon_i -
\mu)
\delta_{il}-t_{il}~~~~~~~~~~~-\Delta_{il}^{\downarrow
\uparrow}~~~~~~~~~~~~-\Delta_{il}^{\downarrow
\uparrow} \\
-\Delta_{il}^{\uparrow \uparrow
*}~~~~~~~~-\Delta_{il}^{\downarrow
\uparrow *}~~~~~~~(-\varepsilon_i + \mu) \delta_{il}
+t_{il}~~~~~0 \\
-\Delta_{il}^{\uparrow \downarrow
*}~~~~~~~~-\Delta_{il}^{\downarrow
\downarrow *}~~~~~~~0~~~~~(-\varepsilon_i +
\mu) \delta_{il}
+t_{il} \end{array}
  \right)
 \left( \begin{array}{c} u_{l \uparrow}
\\
u_{l \downarrow}
\\
v_{l \uparrow}
\\
v_{l \downarrow} \end{array}
\right) =
E \left( \begin{array}{c}
u_{i \uparrow}
\\
u_{i \downarrow  }
\\
v_{i \uparrow}
\\
v_{i \downarrow}
 \end{array} \right),
\end{equation}
while the analogue of the equation of motion for the Green function in Eq. \ref{eq3}
should
be  rewritten as:
\begin{equation}
\label{eq5}
\sum_l \left( \begin{array}{c} \left[(\imath \omega_n -\varepsilon_i + \mu)
\delta_{il}+t_{il}\right] \mb 1~~~~~ \mb \Delta_{il} \\
 \mb \Delta_{il}^+~~~~~ \left[(\imath \omega_n +\varepsilon_i - \mu) \delta_{il}
-t_{il} \right] \mb 1  \end{array} \right)
 \left( \begin{array}{c} \mb G_{11}(l,j;\imath \omega_n)~~
\mb G_{12}(l,j;\imath \omega_n) \\
\mb G_{21}(l,j;\imath \omega_n)~~ \mb G_{22}(l,j;\imath \omega_n) \end{array}
\right)
=
\delta_{ij} \mb 1 
\end{equation}
with the spin dependent 4$\times$4  Green function. Each of  its components
$\mb G_{nm}(l,j;\imath
\omega_n)$ is defined as:
\begin{equation}
\label{eq6}
\mb G_{nm}(l,j;\imath \omega_{\nu})= \left(\begin{array}{cc} G_{nm}^{\uparrow
\uparrow}(l,j;\imath \omega_{\nu}) &
G_{nm}^{\uparrow 
\downarrow}(l,j;\imath \omega_{\nu}) \\ G_{nm}^{\downarrow 
\uparrow}(l,j;\imath \omega_{\nu}) &
G_{nm}^{\downarrow 
\downarrow}(l,j;\imath
\omega_{\nu})
\end{array} \right),~~~~~n,m=1,2~.
\end{equation}
The order parameter for triplet paring reads:
\begin{equation}
\label{eq7}
\mb \Delta_{ij} =
\left(
\begin{array}{cc}
\Delta_{ij}^{\uparrow \uparrow} & \Delta_{ij}^{\uparrow \downarrow} \\
\Delta_{ij}^{\downarrow \uparrow} & \Delta_{ij}^{\downarrow \downarrow}
\end{array} 
\right).
\end{equation}

We assume that the hopping  integrals $t_{ij}$ can take a nonzero values
for
the  nearest
and
next nearest neighbours. For a clean system $t_{ij}$ can be expressed
in
$\mb k$-space by  Fourier 
transform:
$\sub{\epsilon}{k}= \sum_j t_{ij} {\rm e}^{-\imath \mbs R_{ij} \mbs k} $
as:
\begin{equation}
\label{eq8}
\sub{\epsilon}{k} = -2 t (\cos k_x a+
\cos k_ya)+4t'\cos k_xa \cos k_ya,  
\end{equation} 
where $t$ represents the nearest neighbour site amplitude of electron
hopping, while $t'$ corresponds to next nearest neighbour hopping, $a$ denotes the lattice
constant, $\mu$ 
is the 
chemical potential equal to Fermi energy in zero temperature ($\mu = E_F$ for
$T=0$). 
We shall refer to the Greens function matrix  
as   $\mb G (i,j;\imath \omega_n)$ which will be of 2$\times$2 or 4$\times$4
size for 
singlet  or triplet solution. The above
equations have to be 
completed by the self-consistency condition for pairing potential:
\begin{equation}
\label{eq9}
\Delta_{ij}= U_{ij} \frac{1}{\beta} \sum_n {\rm e}^{\imath \omega_n
\eta} 
G_{12} (i,j;\imath \omega_n),
\end{equation}
for singlet pairing case and 
\begin{equation}
\label{eq10}
\Delta_{ij}^{\alpha \alpha'} = U_{ij} \frac{1}{\beta} \sum_n {\rm e}^{\imath
\omega_n
\eta}
G_{12}^{\alpha \alpha'} (i,j;\imath \omega_n)
,~~~~~ \alpha, \alpha'
=\uparrow, \downarrow
\end{equation}
for triplet one, where
$\eta$    is a positive infinitesimal, $\beta=\frac{1}{Tk_B}$ is the
inverse of
temperature $T$ and Boltzman constant $k_B$ (in the units we use here $k_B =1$).  
To simplify 
matters we 
have assumed that the Hartree term $U_{ij}<n_{j-\sigma}>$ can be  absorbed
into the hopping 
integral  $t_{ij}$  and dropped it from  Eqs. (\ref{eq2},\ref{eq3}) and
 (\ref{eq4},\ref{eq5}). As
usual
Eqs.
(\ref{eq9}) and (\ref{eq10}) are to be solved 
together with the corresponding equations for the chemical potential
$\mu$ that
satisfy the
following:
\begin{equation}
\label{eq11}
n= \frac{2}{\beta} \sum_n {\rm e}^{\imath \omega_n \eta} 
G_{11}(i,i;\imath \omega_n)
\end{equation}
for singles or
\begin{equation}
\label{eq12}
n= \frac{2}{\beta} \sum_n {\rm e}^{\imath
\omega_n \eta}
G_{11}^{\uparrow \uparrow}(i,i;\imath \omega_n)  
\end{equation}
for triplets,  
where $n$ is the number of electrons per unit cell.

Here we do not wish to  be very specific about the physical 
nature of the point defects represented by the site energies $\varepsilon_i$. 
We are rather going to  provide a reliable analysis of the simplest 
possible non trivial model.   Thus we take them to be independent
random 
variables defined to have values  $\frac{1}{2} \delta$ and  
$-\frac{1}{2} \delta$ with equal probability, $\frac{1}{2}$, on every site
\cite{Lit98,Mar99}.
As might be expected we shall be interested in the average of 
$\mb G(i,j;\imath \omega_n)$ over the above
ensemble. To 
calculate $\overline {\mb  G} (i,j;\imath \omega_n)$ we
shall 
make use of the Coherent Potential Approximation (CPA) which is  
the best method at hand for the mean field theory of disorder \cite{Ell74}.

\subsection{Van Hove singularities in clean superconductors}

Let's assume that the sites 
form a square lattice. Then for clean system ($\varepsilon_i=0$ for all 
$i$), in the normal
state, where $\Delta_{ij}=0$, the spectrum is given by
$\sub{\epsilon}{k}$ (Eq. \ref{eq8}). It has a
saddle point Van Hove 
singularity at $E_v=4t'$, resulting in the logarithmic divergence of the
density of states
$N(E) \sim -{\rm ln}(E-E_v')$ (Fig. 1b) 
The density of states for a normal state $N(E)$ is defined as:
\begin{equation}
\label{eq13}
N(E)=\frac{1}{N} \sum_{\mbs k} \delta
(E-\sub{\epsilon}{k})
=  \frac{a^2}{4 \pi^2}\int_{ E=\sub{\epsilon}{k}} {\rm d} f \left|
\sub{\nabla}{k}
\sub{\epsilon}{k} \right|^{-1},
\end{equation}
where ${\rm d} f$ is an element of Fermi surface (Fig. 2b)
$N(E)$ reaches its maximum if
Fermi surface  satisfy the relation
$E-\sub{\epsilon}{k}=0$. 
Finally, density of states $N(E)$ can be also expressed by the elliptic
function
of a first kind ${\rm K}(E)$ \cite{Que99}:
\begin{equation}
\label{eq14}
N(E)=\frac{1}{N} \sum_{\mbs k} 
{\rm Im} \frac{1} {E+\imath \eta -\sub{\epsilon}{k}}=
\frac{1}{ 2 \pi^2 t \sqrt{1+ \frac{Et'}{t^2} } } {\rm K}
\left( \sqrt{\frac{16 t^2 - (E-4t')^2}{16t^2(1+\frac{Et'}{t^2})}} \right).
\end{equation}
Van Hove singularities in density of states
(Eq. \ref{eq13},\ref{eq14}) correspond to three characteristic flat
regions of
$\sub{\epsilon}{k}$, where 
$\sub{\nabla}{k}
\sub{\epsilon}{k}=0$.
In Fig. 2a 
we have plotted band energy  $\sub{\epsilon}{k}$ (Eq. \ref{eq8}).
For cuprates the value of next nearest neighbour hopping term $t'$ is
usually
considered as $0< t'< 0.5t$. Here we have chosen $t'=0.45t$ (Eq.
\ref{eq8}).
One can easily
determine 
Van Hove singularities for saddle points: $(|k_x|,|k_y|) =$
$(\pi/a, 0)$, $(0,\pi/a)$  and these corresponding to band edges:
the bottom $(0, 0)$  as well as  top one $(\pi/a,\pi/a)$. 
Three isoenergetic
contours '3', '2', '1' have been 
marked in  Fig. 2a  for $E/t=2$, $1.8$, $1.6$.
They correspond to Fermi surfaces (Fig. 2b) 
for three values of band
filling
$n =0.338$, $0.467$, $0.584$ respectively.
Note that for $n=0.584$, the Fermi surface has
hole like
characteristics
while for $n=0.388$, it corresponds to electron like system.  
For $n=0.467$ the Fermi energy $E_F=1.8t$ passes through the Van Hove
saddle
point singularity.

For the 'on-site' attraction (negative $U$)  $U_{ii}=U$ the linearized gap
equation
(Eq. \ref{eq9})
at
$T_c$ can be
written as:
\begin{equation}
\label{eq15}
1 = \frac{U}{\pi}\int_{-\infty}^{\infty} {\rm d} E \frac{1}{N} \sum_{\mbs
k}
\frac{\delta(E-\sub{\epsilon}{k}-\mu)}{2 E}~{\rm tanh} \left ( 
\frac{\beta_c
E}{2} \right) 
= \frac{U}{\pi} \int_{-\infty}^{\infty} {\rm d} E
\frac{N(E)}{2E}~{\rm tanh} \left( \frac{\beta_c
E}{2} \right),   
\end{equation}
where $\beta_c=1/(T_c k_B)$, and $T_c$ is a critical temperature.

If the interaction is off diagonal  then the Fourier transform of
$U_{ij}$ 
leads to following expression:
\begin{equation}
\label{eq16}
U(\mb k - \mb q)= -|U| \left( \frac{\sub{\eta}{k} \sub{\eta}{q} +
\sub{\gamma}{k} \sub{\gamma}{q}}{4} + 2 \sin k_xa \sin q_xa +
2 \sin k_ya \sin q_ya \right),
\end{equation}
where
\begin{eqnarray}
\sub{\gamma}{k} &=& 2(\cos k_xa + \cos k_ya), \nonumber \\
\label{eq17}
\sub{\eta}{k} &=& 2(\cos k_xa - \cos k_ya). 
\end{eqnarray}
The pairing parameters for corresponding symmetry  of solution: extended
$s$--
$d$-- or $p$--wave one have the following form:
\begin{eqnarray}
\sub{\Delta}{k}^s &=& \Delta_0 \sub{\gamma}{k}, \nonumber \\
\label{eq18}
\sub{\Delta}{k}^d &=& \Delta_0 \sub{\eta}{k}, \\
\sub{\mb \Delta}{k}^p &=& \mb \Delta_0^x \sin(k_x a) +\mb \Delta_0^y
\sin(k_y a), \nonumber
\end{eqnarray}
and $\sub{\mb \Delta}{k}^p$ is a Fourier transform of the matrix ${\mb
\Delta}_{ij}$ (Eq.
\ref{eq7}). Despite of different types possible solutions described by
Eqs.
(\ref{eq3},\ref{eq5},\ref{eq18}) the linearized gap equation for the
critical
temperature can be written in a compact form: 
\begin{equation}
\label{eq19}
1 =
\frac{U}{\pi}
\int_{-\infty}^{\infty} {\rm d} E
~\frac{N_{\alpha}(E)}{2E} ~{\rm tanh} \left( \frac{\beta_c
E}{2} \right),
\end{equation}
depending
on the solution symmetry $\alpha=s,d,p$.

The normal density of states $N(E)=\frac{1}{N} \sum_{\mbs k} \delta ( E
-\sub{\epsilon}{k})$ and the corresponding projected densities
($N_s(E)$,
$N_d(E)$ and $N_p(E)$) 
used in Eq. (\ref{eq19}) can be  expressed in
terms of Green
functions of the normal system as follows: 
\begin{eqnarray}
N(E) &=& -\frac{1}{N} \sum_{\mb k}  
~\frac{1}{\pi} {\rm
Im} G_{11} ( 
\mb k,E) \nonumber, \\
\label{eq20}
N_s(E) &=& -\frac{1}{N} \sum_{\mb k} \frac
{\sub{\gamma}{k}^2}{4} ~\frac{1}{\pi} {\rm
Im} G_{11} 
(\mb k,E) \nonumber,
\\
N_d(E) &=& -\frac{1}{N} \sum_{\mb k} \frac
{\sub{\eta}{k}^2}{4}~\frac{1}{\pi} {\rm
Im}
G_{11}   
(\mb k,E),
 \\
N_p(E) &=&  -\frac{1}{N} \sum_{\mb k} 2 ({\rm sin}
k_x a)^2~\frac{1}{\pi} {\rm
Im}
G_{11}   
(\mb k,E) \nonumber. 
\end{eqnarray}

To calculate the above densities of states we have used the recursion
method described in the Appendix A.
The appropriate densities of states for the clean system $N(E)$,  $N_s(E)$
$N_d(E)$ are plotted in
Fig. 3a. Figure 3b shows the $p$--wave projected density $N_p(E)$ in
comparison to $N(E)$. 

The full line, in Fig. 3a, corresponds to the electron density of states
$N(E)$
with
the spectral dependence of $\sub{\epsilon}{k}$  (Eq. 4) and $t'=0.45$.
The position of the central Van Hove singularity  $E_v=-1.8t$ (Fig. 2)
corresponds
to band
filling
$n=0.467$. Apart from that singularity  one can see another sharp peak in
the 
lower edge of band and much smoother one  
in the upper edge. Both of them are   singularities of
band edges: bottom and top
respectively.
For the 'on site interaction' $U_{ii}$ the shape of $N(E)$ in Fig. 3a and
the
gap equation (Eq. \ref{eq15}) indicates that
the critical temperature $T_c$ should be enhanced effectively
around first two singularities for relatively low electron densities
$n < 1$.
As a matter of fact
such situation can be seen in calculations of $T_c$
(Fig. 4, clean system for curves denoted by '1').
The critical temperature $T_c$ obtained from the Eqs.
(\ref{eq19},\ref{eq20}) for various symmetries of the order parameter
are depicted in Fig. 4a-c by lines denoted by '1' as a function of band filling $n$. 
Figure 4a corresponds to extended $s$--, $d$-- wave cases while
Fig. 4b shows $T_c$
to the $p$--wave
solution. 
In this case there exists a maximum in projected density of states
$N_p(E)$ close to Van Hove singularities in $N(E)$ 
but it is not so sharp as in  $N(E)$ or $N_s(E)$ and $N_d(E)$
because
of the additional smearing term $2 (\sin{k_x a})^2$ in
the formula
for $N_p(E)$ (Eq. \ref{eq20} ). Nevertheless, this secondary peak  also
produces
 the
enhancement of critical temperature $T_c$ (Fig. 4b). 
For comparison, in Fig. 4c we present $T_c$ for
an isotropic 'on-site'  $s$-- wave solution.
Clearly, these calculations support the Van Hove
singularity
scenario. Note that for relatively low temperature Van Hove singularity is
passing 
at $n=0.467$ (Figs. 2, 3a). The lack of the particle-hole symmetry results
in
the shift of
maximum
value of $T_c$ to higher value of electron densities so in the optimal
doping the superconductor is hole superconductor.    
 
Moreover for $n \rightarrow 2$ we observe the degradation of critical
temperature
$T_c$.  
In in this limit of slowly changing density of sates we can apply the
result for
the constant density of states:
\begin{equation}
\label{eq21}
N(E)=\frac{1}{D}~\Theta(E-D/2)~\Theta(-E+D/2),
\end{equation}
where $D$ denotes the bandwidth and  $\Theta(E)$ is the Heaviside step
function.

For small 'on site' attractive interaction $U$, the critical temperature
$T_c^*$ is
given by the analytic formula \cite{Mic90,Mic88}:
\begin{equation}
\label{eq22}
T_c^* = \frac{{\rm e}^{\gamma}}{\pi} \sqrt{n(2-n)}~ {\rm exp} \left(
-\frac{D}{|U|} \right)
\end{equation}

In our case, due to the asymmetry of the electron density of states
$N(E_F)<< 1/D$ for $n  \rightarrow  2$  the degradation of  $T_c$ is
faster $T_c << T_c^*$.

The Van Hove singularities in the spectrum $\sub{\epsilon}{k}$ show up
also in the
projected densities of states $N_s(E)$, $N_d(E)$ (Fig. 3a). Due to the
factors
$\sub{\gamma}{k}$ and $\sub{\eta}{k}$ (Eqs. \ref{eq17},\ref{eq20}) the 
dependences $T_c(n)$
are
severely modified (Fig. 4a, curve '1').
Moreover, the maximum value of  $N_p(E)$ (Fig. 3b) is also close to 
the Van Hove singularity and it results in the optimum
of $T_c$ (Fig. 4b, curve '1').

It is worthwhile to notice here that singularities in the band edges are
important
for an extended $s$--wave case
while the saddle
point one, located in the middle of the band, for 
$d$--wave pairing, similarly to an isotropic $s$--wave case.

The positions of the Van Hove
singularities
result in the strong band filling $n$ dependence of $T_c$ (Fig. 2b).
Again the pairing dominates in selected regions of $n$.
In case of $d$--wave it is the middle region of $n \approx 0.5$ while for
extended
$s$--wave
pairing it is rather low electron densities region ($n \rightarrow 0$).
The high electron
density ( $n \rightarrow 2$) is also possible however the $T_c$ is much
smaller. 
Same relation, as in the 'on-site' 
pairing case,  governs the basic behaviour (Eq.~\ref{eq22}). However
$N(E)=1/D$ should be
substituted by the corresponding projected density of states
$N_{\alpha}(E_F)$.
Thus, away the Van Hove singularity, ($n \rightarrow 2$):
\begin{equation}
\label{eq23}
T_c^{\alpha} \approx~{\rm exp} \left( -\frac{1}{
N_{\alpha}(E_F)
|U|} \right)~~~~\alpha=s,p,d.
\end{equation}
  
In all cases the Van Hove singularities play the mayor role and could be
identified as the
source of a critical temperature $T_c$ raise.
Its dependence on  doping $n$  should be described rather by strongly
changing
function  
contrasting
with the case of a constant density of states (Eq. \ref{eq14}) where $T_c$
is simply $\sqrt{n(2-n)}$. 

However it is also worthwhile to note that depending on the paring
symmetry
different Van Hove
singularities
can
matter and because of non symmetric character of densities of states 
this effect is approximate.  Due to the this asymmetry we observe the
additional shift of the
optimal
doping toward the center of a band.  
Interestingly,  different Van Hove singularities  and corresponding
shifts depend on solution type and  the interaction range. Eventually, an
extend
ed
$s$--wave
solution appears to be an electron superconductor while $d$--wave can be
identified
as hole one. 

The formulae for a critical temperature $T_c$ (Eqs. 
\ref{eq19}) are based on  the integral over the appropriate DOS
which
posses
Van Hove singularities. The effect of shift is stronger for smaller $T_c$
where hyperbolic tangent is smearing the function under integrals. 
It is also worth to note that Van Hove scenario is working better for
superconductors with relatively small transition temperature $T_c$  (which
corresponds to a small
interaction parameter $U$). This can easily be seen  from the  following
function:
\begin{equation}
\label{eq24}
F(T_c,E)=
\frac{2T_c}{(E-\mu)}~{\rm
tanh}
\left( \frac{E-\mu}{2T_c} \right)
\end{equation}
present in the gap equations (\ref{eq9},\ref{eq12}). It can be
interpreted 
as leading to a natural cut-off $E_C$ around the chemical potential $\mu$.
Note 
that if the
temperature $T_c$ is 
small 
then the function $F(T_c,E-\mu)$ is non zero in the narrow range of
energies around $\mu$ only. In fact in the limit $T_c \rightarrow 0$ it 
tends to the Dirac delta function ($F(T_c,E-\mu) \rightarrow
\delta(E-\mu)$)
and the cut-off  is limited to the neighbourhood of   the $E=\mu$ point. 
  Note, that for finite $T_c$, $E_C  \approx 2 T_c$. 
Consequently, for the logarithmic  Van
Hove singularity
in 
the
density of
states near $\mu$ has the form:
\begin{equation}
\label{eq25}
N(E) \approx - N_0~ {\rm ln} \left| \frac{E - \mu}{D} \right|. 
\end{equation}
and we get the Labbe--Bok formula for $T_c$ \cite{Lab87,Bou98,Tsu90}:
\begin{equation}
T_c \sim {\rm exp} \left\{ -\frac{1}{\sqrt{|N_0 U|}}
\right\}.
\label{eq26}
\end{equation}

\section{CPA for disordered Hubbard Model}
As mentioned above, 
the technical question we shall answer in this paper is  what happens 
to above behaviour when the site energies $\varepsilon_i$ are not the same on all sites 
but are
randomly
distributed.
For example in  a binary alloy,  $A_cB_{1-c}$, we have  random
distribution
of site energies: $\varepsilon_i = (\varepsilon_A, \varepsilon_B)$
depending on
occupation at the site $i$ by atom $A$ or $B$ with probabilities $P_{A,(B)}=$
$c$ and $(1-c)$ respectively.
Such problem have been dealt with within the Coherent Potential Approximation (CPA) on a
number
of occasions in the past \cite{Lus73,Ker74,Wys83,Lit92,Lit00a,Mor01,Wen95,Lit98,Mar99,Lit00b}.
Here we shall follow the usual arguments generalized as appropriate. In short, we shall
take
the CPA to mean that 
the coherent potential $\mb \Sigma(E)=\mb \Sigma(i,i;E)$ \cite{Ell74},
in a site
approximation, is defined by the zero value of an averaged  t-matrix $ \mb T(i,i;
E)$. Namely 
\begin{eqnarray}
<\mb T_{\alpha}(i,i; E)> &=&\sum_{\alpha} P_{\alpha} \mb T_{\alpha}(i,i; 
 E) 
\label{eq27}
\\ &=& 
\left< (\mb V_{\alpha}-\mb \Sigma^{\sigma}(E))(\mb 1-[\mb V_{\alpha}-
\mb \Sigma(E)]\overline {\mb G} (i,i; 
 E))^{-1} \right>=0, \nonumber
\end{eqnarray}
where $\alpha=A,B$ specifies the occupation of the site $i$ and hence  the disordered
potential
$\mb V_{\alpha}$.

In case of 'on site attraction' (Eq. \ref{eq1} with $U_{ij}=U_{ii}
\delta_{ij}$). In the limit of small  fluctuations in the paring
potential ($\Delta_{ii}=\Delta_{i}$),  a
constant averaged value
\begin{equation}
\label{eq28}
\Delta_i \rightarrow \overline \Delta ~~~~~{\rm for~all}~i
\end{equation}  
can be applied.

Using now CPA equations  equations (Eqs. \ref{eq27} and
Appendix B) it can be readily shown that due to the disorder $\overline
\Delta$
and $E$ in the clean limit is
renormalized to $\tilde \Delta$
and $\tilde E$  in the same way
 
\begin{eqnarray}
\label{eq29}
 \tilde \Delta (E) &=& \overline \Delta~ \frac{2E- {\rm Tr} \mb \Sigma (E))}{2E} \\
\tilde E &=& E~ \frac{2E-{\rm Tr} \mb \Sigma (E))}{2E} . \nonumber
\end{eqnarray}
On account of the general symmetry  between the  the averaged Green function
$\mb G(i,i;E)$ elements
and  that of the coherent
potential $\mb \Sigma(E)$, for complex energies, it follows that
\begin{equation}
\label{eq30}
G_{11}(i,j;E)=G_{22}(i,j;-E^*),
\end{equation}
\begin{equation}
\label{eq31}
\Sigma_{11}(E)=\Sigma_{22}(-E^*).
\end{equation}
Note that one can write Eq. (\ref{eq29}) in
terms of Matsubara  frequencies ($E=\imath \omega_n$):
\begin{equation}
\label{eq32}
\frac {\tilde \Delta (\omega_n)}{\overline \Delta}= \frac{\tilde
\omega_n}{\omega_n},
\end{equation}
where
\begin{equation}
\label{eq33}
\imath \tilde \omega_n = \imath \omega_n - \Sigma_{11}(\imath \omega_n)
\end{equation}
Eventually  Eq. (\ref{eq31}) leads to the linearized gap equation
(Appendix B):
\begin{equation}
\label{eq34}
1 =  U \int_{-\infty}^{\infty} {\rm d}E~
 \overline{N}(E)
\frac{{\rm
tanh} (\frac{\beta E}{2}) }{2
E}.
\end{equation}
This  equation relates directly to  the similar one of the clean case
(Eq. \ref{eq15}) with the difference due to the
substitution of the
density of states for the pure system $N(E)$
by the averaged one of the doped material $\overline N(E)$.  Evidently,
assuming small fluctuations in the the
pairing potential (Eq. \ref{eq28})  one gets a critical
temperature
$T_c$
weakly depending on disorder.
This is the content of the Anderson theorem
\cite{Mak69,Gyo97} which rests on  the assumption that pairing potential does not
fluctuate
in space $\Delta_i \approx \overline \Delta$ for all $i$. However, it should be  noted
that in short coherence length limit the
situation can be opposite. In that case allowing
the spatial 
fluctuations of the pairing amplitude   
$\Delta_i \neq \Delta_j$ induced by the site energy disorder $\varepsilon_i \neq
\varepsilon_j$ or introduced by  the randomly distributed
attractive centers ($U_{ii} = 0$ for some lattice sites $i$) the Anderson theorem
breaks down even for conventional,  on site, $s$--wave superconductors
\cite{Mor01,Lit00a}.

For a constant pairing parameter as in Eq. (\ref{eq28})  the generic
consequence of
disorder
in the system with on site attraction is the
smearing of
the structure in the averaged density of states $\overline N (E )$ (Eq.
\ref{eq34}). To
illustrate the
consequences  of it in our model we have calculated $\overline N (E )$
using the standard CPA procedure \cite{Lit92,Lit98,Mar99} for $c=0.5$, $\varepsilon_A
=-\frac{\delta}{2}$,  $\varepsilon_B =\frac{\delta}{2}$.

The results  for various values of the scattering strength $\delta$
leading to
the smearing of the Van Hove singularities in the averaged
density of states $\overline{N}(E)$, are shown in
Fig. 5a.
Van Hove singularities are still present here for a relatively
weak disorder strength $\delta \le 1t$ while for stronger one $\delta =
2t$ one can
notice additional splitting of singularities  caused by the model of  
disorder. This is the so called the split band regime \cite{Ell74,Mor01}. Namely the two
peaks
are the
remnant of the Van Hove singularities of the two, A and B, pure metals.

Let us now examine  a disordered system with the inter-site attraction
$U_{ij}$. Here  we assume
that
inter-site pairing parameter $\Delta_{ij}$ can be substituted by its
average $\overline \Delta_{ij}$ \cite{Wen95,Lit98,Mar99}.
Thus, in the case of a singlet
paring (extended $s$-- or $d$-- wave),  the on site impurity potential
$\mb
V_{\alpha}$ can be
expressed as
\begin{equation}
\label{eq35}
\mb V_{\alpha} = \left[\begin{array}{cc} \varepsilon_{\alpha} &
0 \\
0 & -\varepsilon_{\alpha}
\end{array}
\right]~~~~\alpha=A~~ {\rm or}~~~ B
\end{equation}
while the coherent potential matrix has the form: 
\begin{equation}
\label{eq36}
\mb \Sigma(E) = \left[\begin{array}{cc} \Sigma_{11}(E) &
0 \\
0 & \Sigma_{22}(E)
\end{array}
\right].
\end{equation}

Naturally, the averaged Green
function,
$\overline{\mb G}(i,i;E)$ can be expressed as
follows:
\begin{equation}
\label{eq37}
\overline \mb G(i,i; E)= \frac{1}{N} \sum_{k}
\overline \mb G(\mb k; E)= \frac{1}{N} \sum_{k} 
\left[ \begin{array}{cc}
E-\sub{\epsilon}{\mb k}- \Sigma_{11}(E) &  \sub{\overline \Delta}{k}
\\
\sub{\overline \Delta}{k}^{*} &E+\sub{\epsilon}{\mb k}- \Sigma_{22}(E)
\end{array}
\right]^{-1},
\end{equation}

In case of triplet pairing ($p$--wave) instead Eqs.
(\ref{eq35}-\ref{eq37}) the 
following notation should be introduced \cite{Lit00b,Lit01}
\begin{equation}  
\label{eq38}
\mb V_{\alpha} = \left[\begin{array}{cccc} \varepsilon_{\alpha} &
0 & 0 &0 \\
0 & \varepsilon_{\alpha} & 0 & 0 \\
0 & 0 & -\varepsilon_{\alpha} &
0  \\ 0 & 0 &
0 & -\varepsilon_{\alpha} 
\end{array}
\right]~~~~\alpha=A~~ {\rm or}~~ B
\end{equation}
for the impurity potential and the coherent potential
\begin{equation}
\label{eq39}
\mb \Sigma(E) = \left[\begin{array}{cccc} \Sigma_{11}(E) &
0 & 0 & 0\\
0 & \Sigma_{11}(E) & 0 & 0 \\
0 & 0 & \Sigma_{22}(E) & 0 \\
0 & 0 & 0 & \Sigma_{22}(E) 
\end{array}   
\right],
\end{equation}
respectively.
Again the averaged Green function is given by
\begin{equation}
\label{eq40} 
\overline \mb G(i,i; E)= \frac{1}{N} \sum_{\mb k}
\overline \mb G(\mb k; E)= \frac{1}{N} \sum_{\mb k}
\left[ \begin{array}{cc}
(E-\sub{\epsilon}{k}- \Sigma_{11}(E)) \mb 1 &  \sub{\overline \mb \Delta}{k}
\\
\sub{\overline \mb \Delta}{k}^{*} &(E+\sub{\epsilon}{k}- \Sigma_{22}(E)) \mb 1
\end{array}
\right]^{-1},
\end{equation}
while 
the conditionally averaged Green function at the impurity site ($\alpha=A$ 
or $B$) has the following form:
\begin{equation}
\label{eq41}
\mb G_{\alpha}(i,i;E)= \overline \mb G(i,i;E)(\mb 1-[\mb V_{\alpha  }-
\mb \Sigma_{\alpha}(E)] \overline \mb G(i,i;E))^{-1}.
\end{equation}
The averaged paring parameters can be written as in (Eqs. \ref{eq18}):
\begin{eqnarray}
\sub{\overline \Delta}{k}^s &=& \overline \Delta_0 \sub{\gamma}{k},
\nonumber \\
\label{eq42}  
\sub{\overline \Delta}{k}^d &=& \overline \Delta_0 \sub{\eta}{k}, \\
\sub{\overline \mb \Delta}{k}^p &=& \overline {\mb \Delta}_0^x \sin(k_x a)
+\overline {\mb \Delta}_0^y
\sin(k_y a). \nonumber
\end{eqnarray}

For anisotropic $s$--, $d$-- and $p$--wave pairing symmetries the gap
equations
(Eqs. \ref{eq9} and \ref{eq10}) take the form:

\begin{equation}
\label{eq43}
\sub{\Delta}{k} = \frac{1}{N} \sum_{\mb q} \frac{\sub{U}{k-q}}{\pi}
\int^{\infty}_{- \infty} {\rm d} E~ 
{\rm Im} \overline G_{12} (\mb k;E) \frac{1}{{\rm e}^{\beta \omega}+1}
\end{equation}
 for singlets (extended $s$-- and $d$-- wave) and  
\begin{equation}
\label{eq44}
\sub{\mb \Delta}{k} = \frac{1}{N} \sum_{\mb q} \frac{\sub{U}{k-q}}{\pi}
\int^{\infty}_{- \infty} {\rm d} E
~{\rm Im}  \overline \mb G_{12} (\mb k;E) \frac{1}{{\rm
e}^{\beta
\omega}+1}    
\end{equation}
for triplets ($p$--wave) cases.

From Eqs. (\ref{eq37}) and  (\ref{eq40}) it follows
that  off diagonal elements of the 
Green function are 
\begin{equation}
\label{eq45}
\overline G_{12} (\mb k;E) =  \frac{\overline G_{11}
(\mb k ;E) +
\overline G_{22} (\mb k;E)}{2 E -\Sigma_{11}(E) -\Sigma_{22}(E)} \overline
\Delta_0 \sub{\zeta}{k},
\end{equation}
for singlets and
\begin{equation}
\label{eq46}
\overline \mb G_{12} (\mb k;E) =  \frac{\overline \mb G_{11}
(\mb k ;E) +
\overline \mb G_{22} (\mb k;E)}{2 E -\Sigma_{11}(E) -\Sigma_{22}(E)}
( \overline
\Delta_0^x \sin{k_xa}+\overline \Delta_0^y \sin{k_ya}),
\end{equation}   
for triplets,
where $\sub{\zeta}{k}= \sub{\gamma}{k}$ or $\sub{\eta}{k}$.

Interestingly, the linearized gap equation can be written as in a  way  similar to that 
for the
clean system (Eq. \ref{eq19}):
\begin{equation}
\label{eq47}
1= \frac{|U|}{\pi} \int^{\infty}_{- \infty} {\rm d} E~ {\rm tanh}
\frac{E \beta_c}{2}~
{\rm Im} \frac{ \overline G^{s,d,p} (E)}{ 2 E - {\rm Tr} \mb \Sigma
(E)}.
\end{equation}
where  the imaginary parts of  $\overline G^{s,d,p} (E)$  define
the projected densities of states for a
disordered system $\overline{N}_s(E)$, $\overline{N}_d(E)$  and 
$\overline{N}_p(E)$ discussed in
Appendix A.
Moreover, as  in
case of clean system (Eqs.
\ref{eq20}):
\begin{equation}
\label{eq48}
\overline N_{s,d,p}(E)= - \frac{1}{\pi} {\rm Im} \overline G^{s,d,s} (E) =
-\frac{1}{\pi N}
\sum_{ \small \mb k} \sub{\zeta'}{k} {\rm Im}  \frac{1}
{E - \Sigma_{11}(E) - \sub{\varepsilon}{k} + \mu},
\end{equation}
where $\sub{\zeta'}{k}=  (\gamma_{k})^2/4, (\sub{\eta}{k})^2/2$ or $2
(\sin (k_xa))^2$
depending on the symmetry of solution: extended $s$--, $d$-- and $p$--wave.
Figures 5b--d show the corresponding projected densities of states
of a disordered system (Eq. \ref{eq48}): $\overline{N}_s(E)$,
$\overline{N}_d(E)$ and $\overline{N}_p(E)$. Like
for $\overline N(E)$ (Fig. 5a) for weak
disorder, Van Hove singularities survive.
Thus, it is
clear that the smearing  of  the  Van Hove singularity  in Figs. 6a-d
implies a weakening of the Van Hove enhancement of
the critical temperature $T_c$.

For our simple model of disorder of binary alloy A$_c$B$_{1-c}$ with c=0.5
the coherent potential $\Sigma_{11}(E)$, satisfies
the following CPA equation
\cite{Ell74}:

\begin{equation}
\label{eq49}
\Sigma_{11}(E) = (\frac{1}{2} \delta - \Sigma_{11}(E) )
\overline G_{11} (i,i; E)
(\frac{1}{2} \delta + \Sigma_{11}(E) ).
\end{equation}

In Fig. 6a,b we also show the corresponding self energy $\Sigma (E)$
which in one site CPA  depends only on energy $E$ but not on
the
wave vector
$\mb k$. One can see that for a relatively weak disorder strength $\delta$
the maximum of $|{\rm Im} \Sigma_{11}(E)|$ is exactly at Van Hove
singularity
as in the
Born approximation \cite{Gor83,Lit98,Mar99}:
\begin{equation}
\label{eq50}
{\rm Im} \Sigma(E) \approx -\frac{\pi \delta^2}{4} \overline{N}(E).
\end{equation}
The above result is valid also below the critical temperature $T_c$ 
because for off diagonal pairing $\Delta_{ii}=0$ \cite{Lit98,Mar99}
and (Eq. \ref{eq49}) is still valid.

In case with a larger disorder strength $\delta$ the maximum of
 $|{\rm Im} \Sigma_{11}(E)|$ is located in some other
energy  according to the model of disorder we use (Eqs. \ref{eq1},
\ref{eq35}-\ref{eq41}).
Simultaneously the real part of self energy ${\rm Re} \Sigma_{11}(E)$ is
changing with energy $E$ renormalizing the chemical potential $\mu$ (Eq.
\ref{eq48}).  

Let us investigate the additional effect of disorder visible in Eq.
(\ref{eq47}).
Comparing this equation with the clean system one (Eq. \ref{eq19}) one can
notice the difference in the denominator, where in case of disordered
system there is an additional strong scattering term ${\rm Tr}~ \mb \Sigma
(E)$.

To examine it further let us rewrite the gap equation
in terms of Matsubara frequencies $\omega_n$:
\begin{equation}
\label{eq51}
1= \frac{|U|}{\beta_c} \sum_{n} {\rm e}^{\imath \omega_n
\eta}
\frac{ \overline G^{s,d,p} (i,i;\imath \omega_n)}{  \imath \omega_n -
\Sigma_{11}
(\imath \omega_n)},
\end{equation}
for $s$-- $d$-- and $p$--wave symmetry.

Now, let us  approximate the normal state self energy $\Sigma_{11} (\imath
\omega_n)$ and 
projected density of states $\overline N_{\alpha}(\imath \omega_n)$ by:
\begin{eqnarray}
\Sigma(\imath \omega_n) &\approx& -\imath |\Sigma_0|{\rm sgn}(\omega_n) 
\label{eq52}\\
\overline N_{\alpha}(\imath \omega_n) &=& N_{\alpha}(\imath \omega_n -
\Sigma(\imath \omega_n)) \approx N(\omega_n 
+ \imath |\Sigma_0| {\rm sgn} (\omega_n))   \label{eq53} \\ &=&
-\frac{1}{\pi} {\rm Im} G_{11}^{\alpha} (i,i;\imath \omega_n +  \imath
|\Sigma_0| {\rm
sgn}
(\omega_n))
\nonumber
\end{eqnarray}

In Fig. 7a
we have plotted the densities $N^d(\imath \omega)$ $N^p(\imath \omega)$
$N^s(\imath \omega)$ 
 versus
imaginary part of energy
$\imath
\omega$ for  Fermi energy $E_f$ chosen at Van Hove singularity 
($E_f=E_v=0$).

Note that in that region we can roughly approximate the corresponding
projected densities by
a simple formula
\begin{equation}
\label{eq54}
N_{\alpha}(\imath \omega_n)= \frac{a_{\alpha}}{ \omega_n + b_{\alpha}},   
\end{equation}
where $a_\alpha$ and  $b_{\alpha}$ are constants depending on the pairing
symmetry $\alpha= s,d,p$. 
Introducing Eq. (\ref{eq54}) and Eqs. (\ref{eq52},\ref{eq53}) into Eq.
(\ref{eq51}) we get:  
\begin{equation}
\label{eq55}
1= |U|a_{\alpha}T_c \sum_{\omega_n > 0}
\frac{2}{\omega_n +
\Sigma_0} \times \frac{1}{\omega_n +b_{\alpha}}.
\end{equation}
Now we have to perform the summation over $\omega_n$ for clean $\Sigma_0
=0$ and disordered cases  $\Sigma_0
\ne 0$.
After some algebra (Appendix C)
we get
the approximate pair--breaking formula:

\begin{equation}
\label{eq56}
\psi \left(\frac{1}{2}\right) -\psi \left( \frac{1}{2} +
\frac{b_{\alpha}}{2 \pi
T_{c0}}\right)=
\psi \left( \frac{1}{2} +
 \frac{\Sigma_0}{2 \pi
T_c} \right) -
\psi \left( \frac{1}{2} +  \frac{\Sigma_0}{2 \pi
T_c} +
\frac{b_{\alpha}}{2 \pi
T_{c}} \right),
\end{equation}

Note that in the limit  the constant density of states  $N_{\alpha}(i \omega)
=$ const.
(Eq. \ref{eq54}), $b_{\alpha} \rightarrow \infty$, we get automatically the  
the standard Abrikosov--Gorkov formula
\cite{Mak69}
with a characteristic pair--breaking parameter $\rho_c=|{\rm
Im}\Sigma_0|/(2 \pi T_{c})$:
\begin{equation}
\label{eq57}
{\rm ln}\left( \frac{T_c}{T_{c0}}\right) = \psi \left(\frac{1}{2}\right) -
\psi \left( \frac{1}{2} + \rho_c \right)
\end{equation}
In Eqs. (\ref{eq56} and \ref{eq57}) $T_{c0}$ denotes  the critical
temperature in a clean system, while $T_c$ is the critical temperature  in
dirty one. 

In Fig. 7b we plot $T_c/T_{c0}$ versus $\Sigma_0/(2 \pi
T_{c0})$ for
few values of
$b_{\alpha}$. Here the chemical potential $\mu$ was fixed at the saddle
point Van
Hove
singularity $E_{v1}$.
Interestingly, the
slop of the curve (Fig. 7b) is increasing with decreasing $b_{\alpha}$
indicating
that in presence of Van Hove
singularities 
  for a disordered system
we get weaker decreasing of $T_c$ than for a flat density of states.
 Thus,
for $d$-wave pairing $b_{\alpha}=0.12$ and
for $p$--wave $b_{\alpha}=0.45$ superconducting phase is
more stable than in case of extended $s$--wave pairing, where $b_{\alpha}
\rightarrow \infty$ (Fig. 7a).   
This is the main result obtained in this section. 
In spite of very rough approximation used here  
the results show that
the Van Hove singularity influences  the  
pair--breaking effect. Namely it makes superconductivity more robust 
increasing the critical strength of disorder 
$\Sigma_0$
needed to break the Cooper pairs.

\section{Critical Temperature for Disordered superconductors}

Let us now turn to the case where both superconductivity and disorder are 
present\cite{Lus73,Ker74,Wys83,Lit92} and calculate the critical temperature
$T_c$ self-consistently.
Although
the full CPA program can be implemented for the 
problem defined by Eqs. (\ref{eq9}-\ref{eq10} and \ref{eqb1}-\ref{eqb4})
\cite{Lit92,Mor01}
and
the specification of
the site 
energy ensemble, it is convenient to make the approximation, valid when
the coherence length $\xi_0$ is much larger then the lattice spacing, that 
the pairing potential $\Delta_{ij}$ does not fluctuate very much and replace 
it in equations (\ref{eq3} and \ref{eq5}) by its 
average value $\overline \Delta_{ij}$
\cite{Gyo97}. 
For conventional isotopic $s$--wave pairing 
the gap equation at $T_c$ takes the simple form (Eq. \ref{eq34}).
Thus, the critical temperature $T_c$ at the optimal 
doping should  be only slightly reduced by disorder due to smearing of the 
density of
states $\overline N(E)$.
The results of numerical calculations for four
different values of a disorder
strength $\delta$ in case of the 'on site' attraction
is presented in Fig. 4c. Clearly, in this case the critical temperature
$T_c$ is 
sightly reduced by the effect of the density of states.

On the other hand for the 'off diagonal' attraction   case 
the linearized gap equation,  in presence of disorder, is given by the formula (Eq.
\ref{eq47}). 
Solving it for extended $s$--  $d$-- and $p$--
wave pairing symmetries
we get  
the critical temperature
$Tc$ versus band filling.
The results for various $\delta$  are presented in Fig.
4a (extended $s$-- and $d$--wave) and 4b ($p$--wave). Here, we
observe significant degradation of $T_c$ in all three anisotropic pairing 
cases. Moreover, in some regions of electron concentration,   disappearing
of a supercondacting phase can be noticed for relatively 
weak disorder ($\delta \le 0.6t$). Especially this happens to the $s$--wave
case with high
electron concentration $n \rightarrow 2$ and $n \approx 0.8$ as well as 
the $d$--wave case for $n \approx 1.2$ (Fig. 4a).  The dramatic
degradation of extended
$s$--wave
superconductivity for $n> 0.3$ can be also seen for stronger disorder
($\delta=1t$). In that region of electron concentration $n$, due to the
binary alloy model of
disorder A$_c$B$_{1-c}$ and $c=0.5$, the position of  
energy
$E$ for the maximum of
a pair--breaking term $|{\rm Im}
\Sigma_{11}(E)|\approx \Sigma_0$  (Fig. 6) coincides with the
chemical
potential
$\mu$ making the pair--breaking mechanism very efficient.
Similar behaviour can be seen for $p$--wave superconductor (Fig. 4b).
Clearly, for large enough $|{\rm Im}
\Sigma_{11}(E)|$ (Fig. 4a--b) the superconductivity is destroyed by the
pair-breaking
effect
shrinking the   region of  $n$  for   $T_c > 0$. 
In the same time the corresponding projected densities are not strongly effected
(Fig. 5).
The results for anisotropic paring in Figs. 4a--b are  contrasting with
Fig. 4c where we
plotted the results for the on site
solution
with $U=U_{ii}$. Here, (Fig. 4c) the region of band filling $n$,  where a
superconducting
solution exists, does not change with disorder at all.

\section{Conclusions and Remarks}

We have analyzed the effect of disorder on disordered Hubbard model with
local and non-local, nearest neighbour, interactions as well as  
nearest and  next
neighbour electron hopping terms.  
We have got numerical and analytical results confirming previous
works on the similar model 
with a simple band energy \cite{Lit98,Mar99} $\sub{\epsilon}{k} = -2 t
(\cos k_x a+\cos k_ya)$. Such dispersion relation, introduces
the electron-hole symmetry for half filled band $n=1$ and  locates the
saddle
point Van Hove
singularity exactly
in the center of the band (Figs. 1a,b). 
Including the additional  hopping $t'$, we break the electron hole
symmetry in
the
densities of
states, resulting in shifting the central Van Hove saddle point
singularity to
the bottom of 
the band. 
In the same time another Van
Hove singularity, located at the bottom band edge, appears to be
important.
Thus the effect of Fermi surface distortion make place both of
singularities very close to each other For some of band filling values
$n \approx 0.4$ both of singularities play important roles. 
It is clear after analyzing
the appropriate projected densities of states $N_{\alpha}$. Moreover
various  pairing symmetries choose different 
Van Hove singularities. $D$-wave symmetry is favored in case of the
regions of 
band fillings $n$ with chemical potential $\mu$ near the saddle point
singularity $E_v$ while 
extended $s$-- wave pairing symmetry is more likely as far as the bottom
edge singularity is concerned.  
Interestingly, on account of the non-symmetricity of electron density of
states 
the Van Hove scenario is 
fulfilled only
approximately. 
Here we observe a small
interaction dependent, shift of
the optimal
doping electron concentration $n \approx n_{op}$ towards the center of the
band. As a result of the above we conclude that  Van Hove singularity is
important for
all discussed
symmetries of order parameter. Interestingly, in presence of an additional
electron hopping to the next
nieghbour
lattice site $t'$, the shift of the saddle point Van Hove singularity in the density of
states explains that  the superconductor with an extended $s$--wave
symmetry
is rather of 
electron type while with $d$--wave symmetry case is of
hole nature.  
Note also that, the positions of  the maximum value of $\overline N_s(E)$,
$\overline N_d(E)$,   $\overline N_p(E)$ (Fig. 5)
are not affected by small disorder. 
In spite of small density of states effects leading to smearing 
peaks in corresponding densities: $\overline N_s(E)$, $\overline N_d(E)$
and
$\overline N_p(E)$ (Eq. \ref{eq48}, Fig. 5),
the critical
temperature $T_c$, plotted in Figs. 4a--b, is degradated strongly with
disorder.  This is due to the pair--breaking term $\Sigma_{11}(\imath
\omega_n)$
(Eq. \ref{eq51}). In the negative $U$, 'on-site',
interaction there is a quite different situation.
Here disorder causes only small
decrease of $T_c$ (Fig. 4c) via a density of states effect (Fig. 5a).
So, the most  interesting effect arises from Eqs. (\ref{eq51},\ref{eq55}),
where $\Sigma_{11}(E)$ acts as a  pair breaker.

Concluding our results we would like to stress that 
the Van Hove singularity does not make the
decrease in $T_c$ more
pronounced as one can expect (because of singularities are present in
the self energy ${\rm Im} \Sigma_{11}(E) \sim N(E) $, Fig. 6). In 
fact singularities make it weaker.
We have analyzed this  effect  very carefully
in Sec. III finding an approximate  pair--breaking formula (Eq. \ref{eq56}).  
Here, the Van Hove singularity influenced  the
pair--breaking Abrikosov--Gorkov curve, changing  its slope  of the $T_c$
versus
$ \Sigma_0$. Similar effect has been observed  in the experimental 
results
for Zn--doped LSCO \cite{Kar00}.
Alternatively, this effect can be  also explained assuming 
anisotropic scattering potentials \cite{Har98,Har00}.

Finally, we observed the dependence of $T_c$ 
on band filling $n$. Our results for
$d$--wave
superconductor show that the Van Hove scenario is valid even in
presence of weak disorder (Figs. 4). 
Similar experimental results were obtained by measuring 
$T_c$, in various cuprate
compounds, as a
function of hole
concentration where Cu were substituted by Zn \cite{Ber96,Tal97}. 
The concentration of Zn
was there the measure of disorder.  

It should be however noted that high
$T_c$ cuprates are strongly correlated electron systems and the mean
field approach basing on  the effective intersite attraction $U_{ij}$,  presented here
has
a limited applicability \cite{Mic97}.
Strictly speaking a more realistic model should posses, beside an
intersite attraction, a
strong Coulomb repulsion term.
Although the approximations we used in the present paper aimed to explain  the
effect of the Van Hove singularity in a presence of disorder in a weak coupling 
regime, the preliminary calculations using simultaneously slave
boson technique and CPA \cite{Kra01} seams to support the general
arguments of the Van Hove  singularities significance for
superconducting cuprates
conjectured
here.

\setcounter{section}{0}
\def\thesection{A} 
\def\theequation{A.\arabic{equation}}  

\section{Appendix}
\setcounter{equation}{0}
\def\theequation{A.\arabic{equation}}  

In this Appendix we apply  the recursion method to calculate appropriate 
densities of
states. 
Let us investigate the projected densities if states  $N_s(E)$, $N_d(E)$, 
$N_p(E)$ and the
corresponding Green functions: $\overline G^s(E)$, $\overline G^d(E)$
and $\overline G^p(E)$:
\begin{eqnarray}
N_s(E) &=& -\frac{1}{\pi}  {\rm Im} G^s(E) = -\frac{1}{N} \sum_{\mb k}
\frac
{\sub{\gamma}{k}^2}{4} ~\frac{1}{\pi} {\rm
Im} G_{11} 
(\mb k,E) \nonumber,
\\
N_d(E) &=&  -\frac{1}{\pi}  {\rm Im} G^d(E) =-\frac{1}{N} \sum_{\mb k}
\frac
{\sub{\eta}{k}^2}{4}~ \frac{1}{\pi} {\rm
Im}
G_{11}   
(\mb k,E), \label{eqa1} \\
N_p(E) &=&  -\frac{1}{\pi}  {\rm Im} G^p(E) =-\frac{1}{N} \sum_{\mb k}
2 (\sin k_x)^2~ \frac{1}{\pi} {\rm
Im}
G_{11}
(\mb k,E), \nonumber
\end{eqnarray}
where $\sub{\gamma}{k}$ and $\sub{\eta}{k}$ were defined in Eq. 11.

Noting trigonometric identities:
\begin{eqnarray}
\frac{\sub{\gamma}{k}^2}{4} &=& \frac{1}{2}(\cos 2 k_x  + \cos 2 k_y) +1
+  2 \cos k_x  \cos k_y, \nonumber \\
\frac{\sub{\eta}{k}^2}{4} &=& \frac{1}{2}(\cos 2 k_x  + \cos 2 k_y) +1
-  2 \cos k_x  \cos k_y, \label{eqa2} \\
2 (\sin k_x)^2 &=& 1- \cos2 k_x, \nonumber   
\end{eqnarray}
the Green functions $G^s(E)$, $G^d(E)$ and  $G^p(E)$ can be easily found
as a
combination of diagonal 
and off diagonal
Green functions 
$G_{11}(i+\delta i,i +\delta j,E)=G_{\delta i, \delta j}(E)$, see the
notation in the schematic picture (Fig. 8):
\begin{eqnarray}
G^s(E) &=& G_{00}(E)+G_{20}(E)+2G_{11}(E), \nonumber \\
G^d(E) &=& G_{00}(E)+G_{20}(E)-2G_{11}(E), \label{eqa3} \\
G^p(E) &=& G_{00}(E)-G_{20}(E). \nonumber
\end{eqnarray}
Function $G_{00}(E)$, $G_{20}(E)$ and $G_{11}(E)$ have been calculated
using the recursion method \cite{Lit95,Lit98,Mar99}.

The above procedure can be also used to calculate the average of projected
densities  $\overline N_s(E)$, $\overline N_d(E)$ and $\overline N_p(E)$as
well as
averaged Green functions $\overline G^s(E)$,  $\overline G^d(E)$ and
$\overline G^p(E)$ can 
be calculated via substitution $E$ by $E -\Sigma_{11}(E)$, where
local $\Sigma_{11}(E)=\Sigma_{11}(i,i,E)$ should be found  
self-consistently according to CPA conditions (Eqs. \ref{eq37}-\ref{eq43}):
\begin{eqnarray}
\overline G^{s,d,p}(E)= G^{s,d,p}(E-\Sigma_{11}(E)) \nonumber \\
\overline N_{s,d,p}(E)= N_{s,d,p}(E-\Sigma_{11}(E)).
\end{eqnarray}

\setcounter{section}{0}
\def\thesection{B}
\def\theequation{B.\arabic{equation}}  

\section{Appendix}
\setcounter{equation}{0}
\def\theequation{B.\arabic{equation}}  

In this  Appendix we apply CPA equations (Eqs. \ref{eq18}-\ref{eq21}) to
the disordered
Hubbard 
model with the 'on site attraction' Eq. \ref{eq1} ($U_{ij} =U_{ii}
\delta_{ij}$) and discuss the Anderson
theorem \cite{And59,Gyo97}. 
The averaged Green
function,
$\overline{\mb G}(i,i;E)$ can be expressed as
follows:  
\begin{equation}
\label{eqb1}
\overline \mb G(i,i; E)= \frac{1}{N} \sum_{k}
\overline \mb G(\mb k; E)= \frac{1}{N} \sum_{k}
\left[ \begin{array}{cc}
E-\sub{\epsilon}{k}- \Sigma_{11}(E) &  -\Sigma_{12}(E)
\\
-\Sigma_{21}(E) &E+\sub{\epsilon}{k}- \Sigma_{22}(E)
\end{array}
\right]^{-1},
\end{equation}
while conditionally averaged Green function has the form
\begin{equation}
\label{eqb2}
\mb G_{\alpha}(i,i;E)= \overline \mb G(i,i;E)(\mb 1-[\mb V_{\alpha  }-
\mb \Sigma_{\alpha}(E)] \overline \mb G(i,i;E))^{-1}.
\end{equation}

The disordered potential in Eq. \ref{eq18} has the following form
\cite{Lit92,Lit00a,Mor01}:
\begin{equation}
\label{eqb3}
\mb V_{\alpha} = \left[\begin{array}{cc} \varepsilon_{\alpha} &
-\Delta_{\alpha} \\
-\Delta_{\alpha}^* & -\varepsilon_{\alpha}
\end{array}
\right],
\end{equation}
where $\varepsilon_{\alpha}= \varepsilon_A$ or
$\varepsilon_B$ corresponds to
different site energies of the lattice site while
$\Delta_{\alpha}= \Delta_A$ or $\Delta_B$ relates to the different paring
potential
in an alloy $A_cB_{1-c}$.

Clearly, the coherent potential can be written as
\begin{equation}
\label{eqb4}
\mb \Sigma(E) = \left[\begin{array}{cc} \Sigma_{11}(E) &
\Sigma_{12}(E) \\
\Sigma_{21}(E) & \Sigma_{22}(E)
\end{array}
\right].
\end{equation}

Function $\overline G_{12}(i,i;E)$  can be obtained from  Eq.
\ref{eqb1}

\begin{equation}
\label{eqb5}
\overline G_{12}(i,i;E)= 
 \frac{ \overline G_{11}(i,i;E)
+
\overline G_{22}(i,i;E)}{2 E - \Sigma_{11}(E) -
\Sigma_{22}(E)}
 \Sigma_{12}(E)=
\left< \frac{ G_{11}^{\alpha}(i,i;E)
+
G^{\alpha}_{22}(i,i;E)}{2 E - \Sigma_{11}(E) -
\Sigma_{22}(E)}
 \Sigma_{12}(E) \right>
\end{equation}

While from Eq. \ref{eqb4} we have the following relations:
\begin{equation}
\label{eqb6}
{\rm Tr} \mb G^{\alpha}(i,i;E) =
{\rm Det} \mb G^{\alpha}(i,i;E) \left(\frac{  {\rm Tr} \overline  \mb
G(i,i;E)}{ {\rm Det} \mb
G(i,i;E)}  - \Sigma_{11}(E) - \Sigma_{22}(E)\right) 
\end{equation}
 \begin{equation}
\label{eqb7}
G_{12}^{\alpha}(i,i;E) =
{\rm Det} \mb G^{\alpha}(i,i;E) \left(\frac{  \overline  
G_{12}(i,i;E)}{ {\rm Det} \mb
G(i,i;E)}  - \Delta_{\alpha} - \Sigma_{12}(E)\right)
\end{equation}

substitution of Eq. \ref{eqb6} and \ref{eqb7} into the right and
left hand
sides of Eq. \ref{eqb5} respectively  an 
equation on $\Sigma_{12}(E)$
and $\Delta_{\alpha}$:

\begin{equation}
\label{eqb8}
\frac{2 E~ \Sigma_{12}(E)}{2 E - \Sigma_{11}(E) -
\Sigma_{22}(E)}
=  \frac{ \left<\Delta_{\alpha} {\rm Det} \mb G^{\alpha}(i,i;E)
\right>}{ \left<
{\rm Det}
\overline \mb G(i,i;E) \right>}
\end{equation}

Using \ref{eqb6} and factorizing of above we get:
\begin{equation}
\label{eqb9}
\frac{2 E~ \Sigma_{12}(E)}{2 E - \Sigma_{11}(E) -
\Sigma_{22}(E)}  \approx \frac {c {\rm Tr} \mb G^A(i,i;E)
\Delta^A
+(1-c) {\rm Tr}
\mb G^B(i,i;E) \Delta^B}{ {\rm Tr}
\overline \mb G(i,i;E)}
\end{equation}
In the limit of small  paring  potential fluctuations it can be used as a
constant $\Delta_i \rightarrow \overline \Delta= {\rm const.}$ to
satisfy
the Anderson
theorem. Then
\begin{equation}
\label{b10}
\frac{2 E~ \Sigma_{12}(E)}{2 E - \Sigma_{11}(E) -
\Sigma_{22}(E)}
\approx  \overline \Delta
\end{equation}
and renormalized quantities of $\omega$ and $\overline \Delta$ reads as:
\begin{eqnarray}
2\tilde E &=& 2 E -  {\rm Tr}
\mb \Sigma (E)
\nonumber \\
\tilde \Delta (E) &=& \Sigma_{12}(E) =  \frac{(2 E -  {\rm
Tr}
\mb \Sigma (E) )}{2 E} \overline \Delta.
\label{eqb11}
\end{eqnarray}
This lead to the same renormalization in the pair potential
$\Delta$ and
energy $E$:
\begin{equation}
\label{eqb12}    
\frac {\tilde \Delta (E)}{\overline \Delta}= \frac{\tilde
E}{E}
\end{equation}

\begin{eqnarray}
\label{eqb13}    
\overline \Delta &=& 
-\frac{U}{\pi} \int_{-\infty}^{\infty} {\rm d} E ~\frac{{\rm Im}
(\overline G_{11}(E) + \overline G_{22}(E))\tilde \Delta}{2
\tilde E} 
\frac{1}{{\rm e}^{\beta E}+1} \\
&=& -\frac{U}{\pi} \int_{-\infty}^{\infty} {\rm d} E ~\frac{{\rm Im}
\overline G_{11}(E) \overline \Delta }{2
E}{\rm
tanh} \left(\frac{\beta E}{2}\right) , \nonumber
\end{eqnarray}
and the linearized gap equation
\begin{equation}
\label{eqb14}  
1 =  U \int_{-\infty}^{\infty} \overline{N}(E) {\rm d} E
\frac{{\rm
tanh} (\beta_c E) }{2
\omega},
\end{equation}
where $N(E)$ denotes normal state DOS: 
\begin{equation}
\overline N(E)= -\frac{1}{\pi} {\rm Im} G_{11} (E) .
\end{equation}

\setcounter{section}{0}
\def\thesection{C}
\def\theequation{C.\arabic{equation}}  
\section{Appendix}
\setcounter{equation}{0}
\def\theequation{C.\arabic{equation}}  
In this Appendix we apply CPA to anisotropic superconductor and
evaluate the approximate
formula of the pair--breaking effect.
Starting from the clean system we assume that the linearized gap equation can
be written (Eqs. \ref{eq53}-\ref{eq57}) by: 
\begin{eqnarray}
\label{eqc1}
1 &=& |U|a^{\alpha}\pi T_{c0} \sum_{\omega_n > 0}
\frac{2}{\omega_n} \frac{1}{\omega_n +b_{\alpha}} \\
 &\approx& \frac{|U|a^{\alpha}2 \pi T_{c0}}{b_{\alpha}} \sum_{\omega_n >
0}^{\omega_n^c}
\left( \frac{1}{\omega_n} -\frac{1}{\omega_n +b_{\alpha}} \right). \nonumber 
\end{eqnarray}
where $\omega_n^c$ is a cut-off in the summation (\ref{eqc1})
As $\omega_n^c$ is very large, it yields
\begin{equation}
\label{eqc2}
\frac{b_{\alpha}}{|U|a_{\alpha}} \approx 
 \psi \left(\frac{1}{2}\right) -
\psi \left( \frac{1}{2} + \frac{b_{\alpha}}{2 \pi
T_{c0}}\right)
\end{equation}
On the other hand  
for disordered system we have:
\begin{eqnarray}
\label{eqc3}
1 &=& |U|a^{\alpha} \pi T_c \sum_{\omega_n > 0}
\frac{2}{\omega_n+ |\Sigma_0| } \times \frac{1}{\omega_n +|\Sigma_0|
+b_{\alpha}} \\
 &\approx& \frac{|U|a^{\alpha}2 \pi T_c}{b_{\alpha}} \sum_{\omega_n >
0}^{\omega_n^c}
\left( \frac{1}{\omega_n+|\Sigma_0|} -\frac{1}{\omega_n +b_{\alpha}+|\Sigma_0|}
\right). \nonumber
\end{eqnarray}
Similarly (to Eqs. \ref{eqc1},\ref{eqc2}) it leads to
\begin{equation}
\label{eqc4}
\frac{b_{\alpha}}{|U|a_{\alpha}} \approx
\psi \left( \frac{1}{2} +  \rho_c \right) -
\psi \left( \frac{1}{2} + \rho_c + \frac{b_{\alpha}}{2 \pi
T_{c}}\right),
\end{equation}
and finally comparing Eqs. (\ref{eqc2} and \ref{eqc4}) we get
\begin{equation}
\label{eqc5}
\psi \left(\frac{1}{2}\right) -\psi \left( \frac{1}{2} +
\frac{b_{\alpha}}{2 \pi
T_{c0}}\right)=
\psi \left( \frac{1}{2} +  \rho_c\right) -
\psi \left( \frac{1}{2} + \rho_c + \frac{b_{\alpha}}{2 \pi
T_{c}}\right),
\end{equation}

where $\rho_c$ is a pair--breaking parameter
\begin{equation}
\label{eqc6}
\rho_c = \frac{|\Sigma_0|}{2 \pi T_{c}}
\end{equation}
and $T_{c0}$ is the critical temperature for a clean superconductor.

\newpage

\section*{ACKNOWLEDGMENTS}
This work has been partially supported by KBN grant No. 2P03B09018.
A part of this work has been done during the stay in  the Abdus Salam
International Centre for Theoretical Physics in Trieste. The author would
like to thank  Prof. K.I. Wysokinski, Prof. B. L. Gy\"{o}rffy and 
Dr. J.F. Annett  for
helpful discussions.

\begin{figure}[htb]
\leavevmode

\vspace{2cm}

\hspace{3.5cm}
\epsfxsize=6.5cm
\epsffile{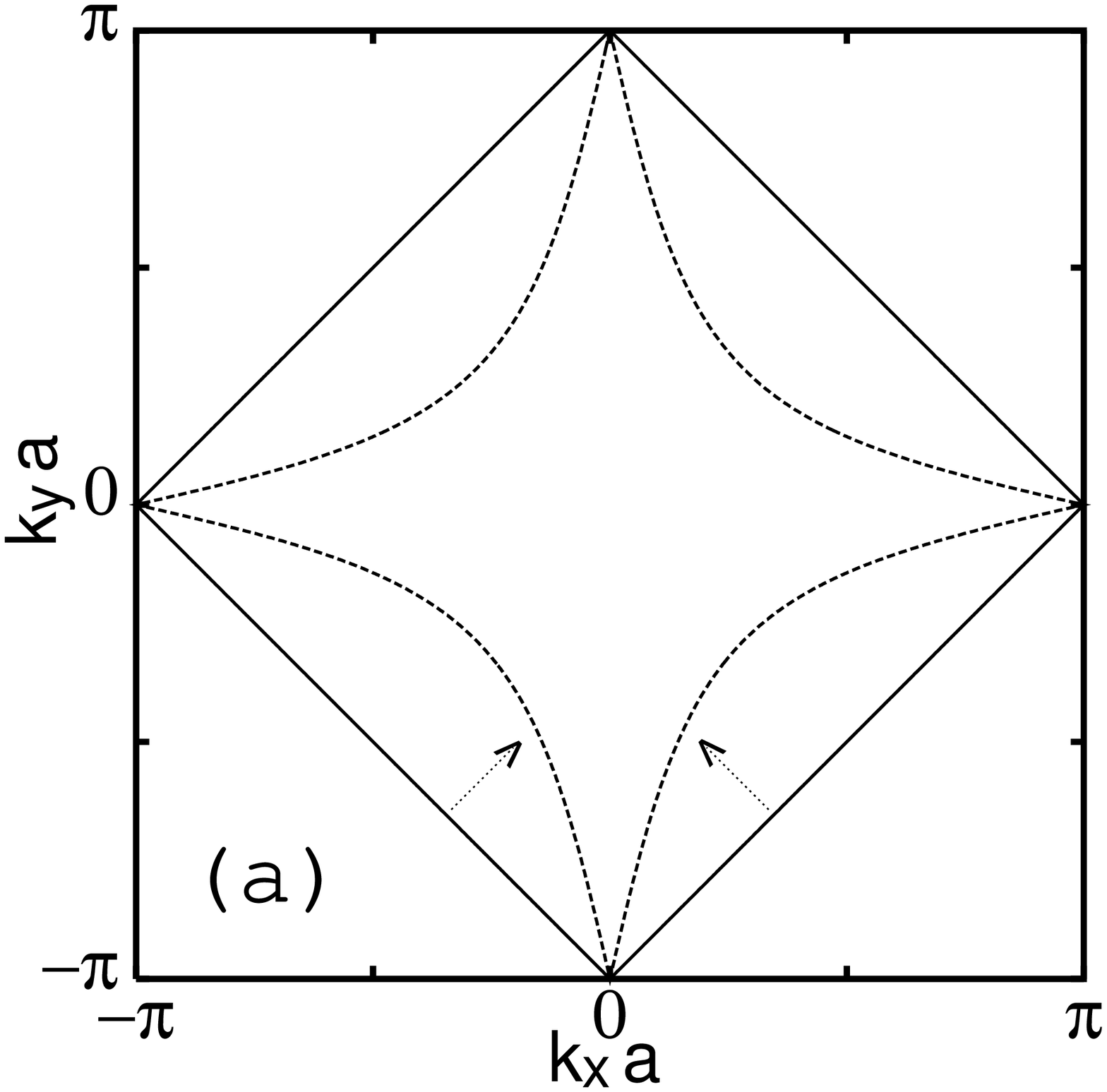}

\vspace{-1.0cm}

\hspace{2.0cm}
\epsfxsize=6.5cm
\epsffile{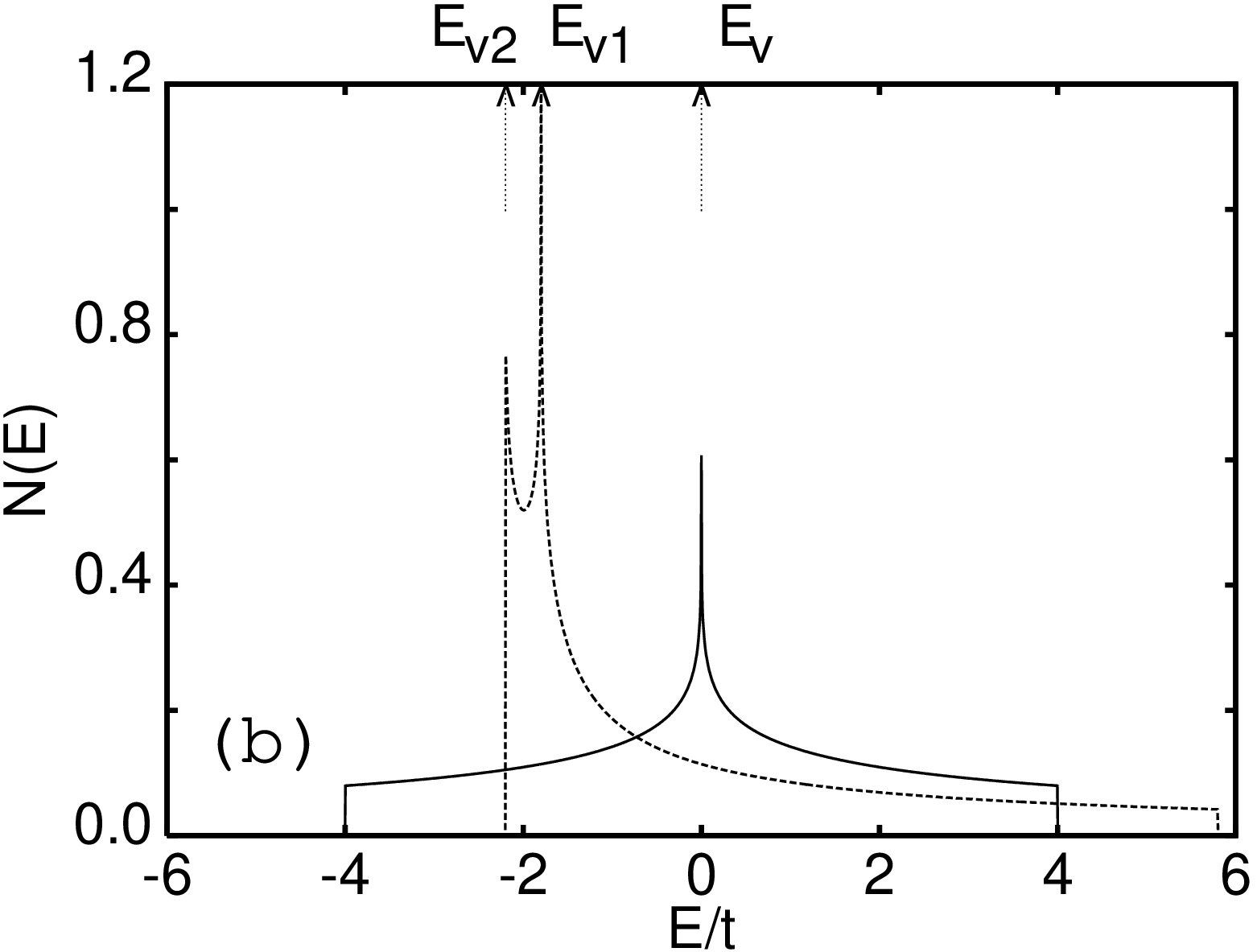}
\caption {(a) The distortion of the Fermi surface by
the
next
neighbour electron hopping $t'$.
The full line denotes Fermi surface for $t'=0$ while the dashed line
$t'=0.45t$.
Fermi  energy is at Van Hove singularity $E_F=E_v$. (b) The electron
density of states $N(E)$ for the 2D lattice, where  the full line
corresponds
to
$t'=0$ and  the dashed line to $t'=0.45t$. Arrows denote Van Hove
singularities $E_v$, $E_{v1}$ and  $E_{v1}$, respectively ($E_F=0$).}
\end{figure}
\vspace{-1.5cm}

\begin{figure}[htb]
\leavevmode

\hspace{2cm}
\epsfxsize=7.5cm
\epsffile{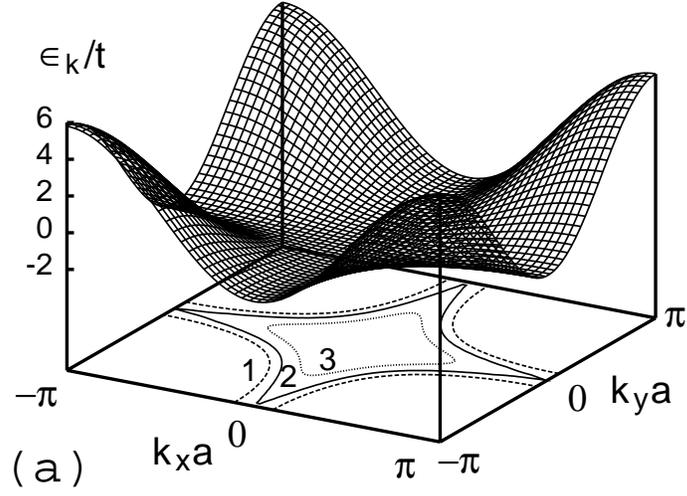}

\vspace{-2.5cm}

\hspace{3cm}
\epsfxsize=7.5cm   
\epsffile{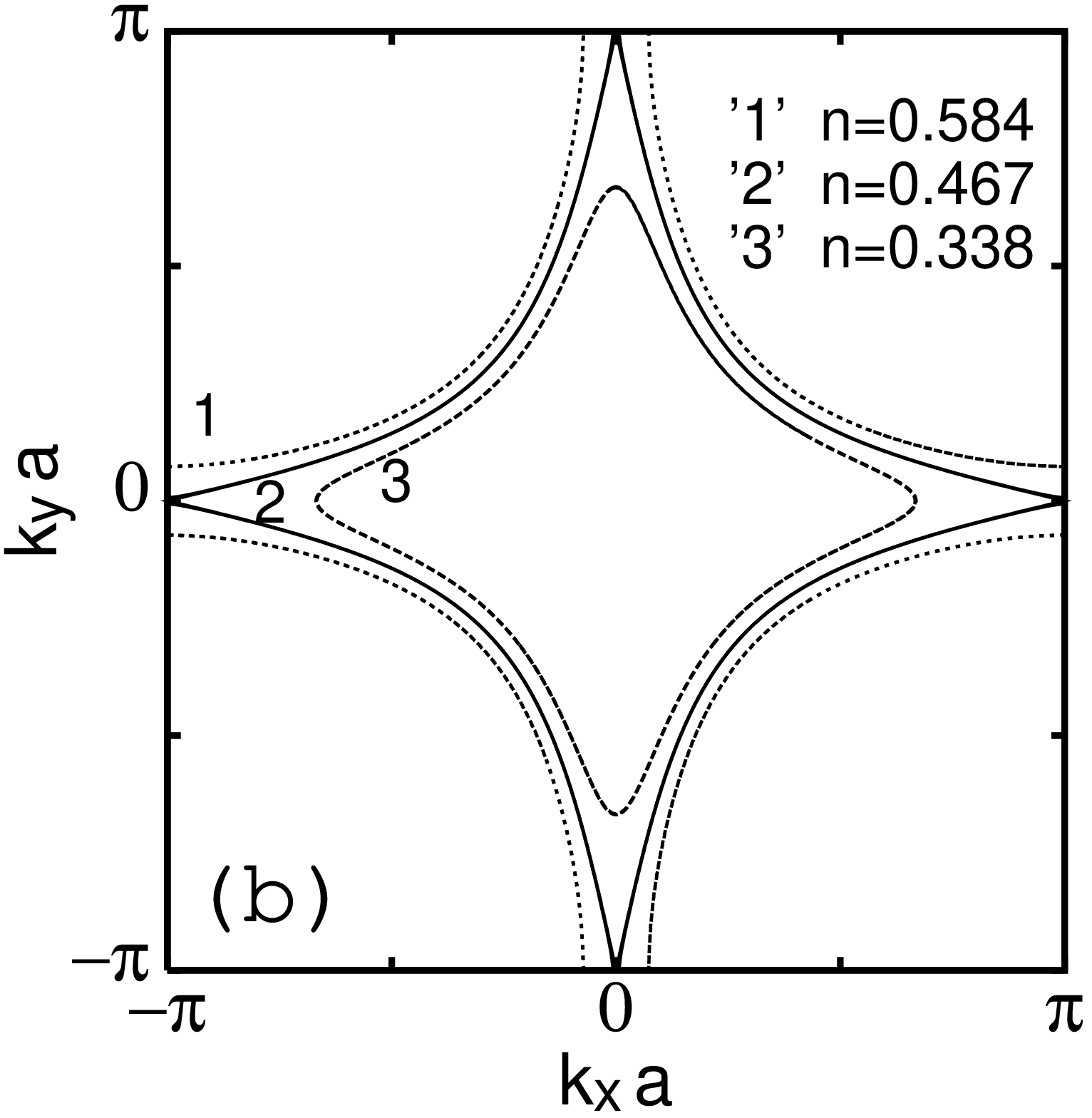}
\vspace{3cm}

\caption{The band structure (a) and
Fermi surfaces (a,b) for the one band electron
structure
with next nearest neighbour hopping: $\sub{\epsilon}{k}  = -2 t
(\cos
k_x +
\cos k_y)+4t'\cos k_x \cos k_y$ with $t'=0.45t$, and three different
band fillings:
$n=0.584$ (1),
$n=0.467$ (2), $n=0.388$ (3).}
\end{figure}

\begin{figure}[htb]
\leavevmode

\epsfxsize=5.0cm
\hspace{2.5cm}
\epsffile{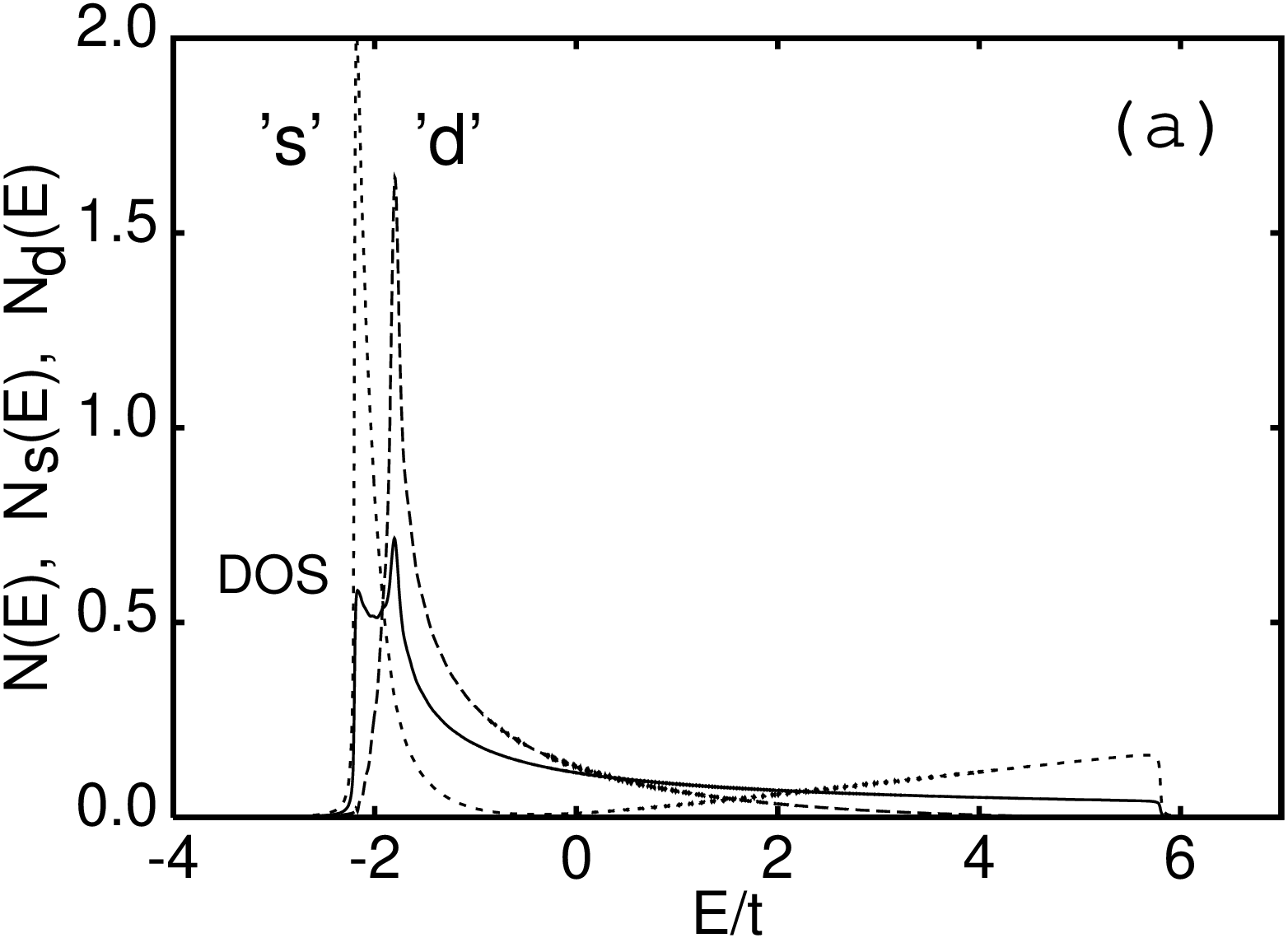}

\vspace*{-1.5cm}

\epsfxsize=5.0cm
\hspace{2.5cm}
\epsffile{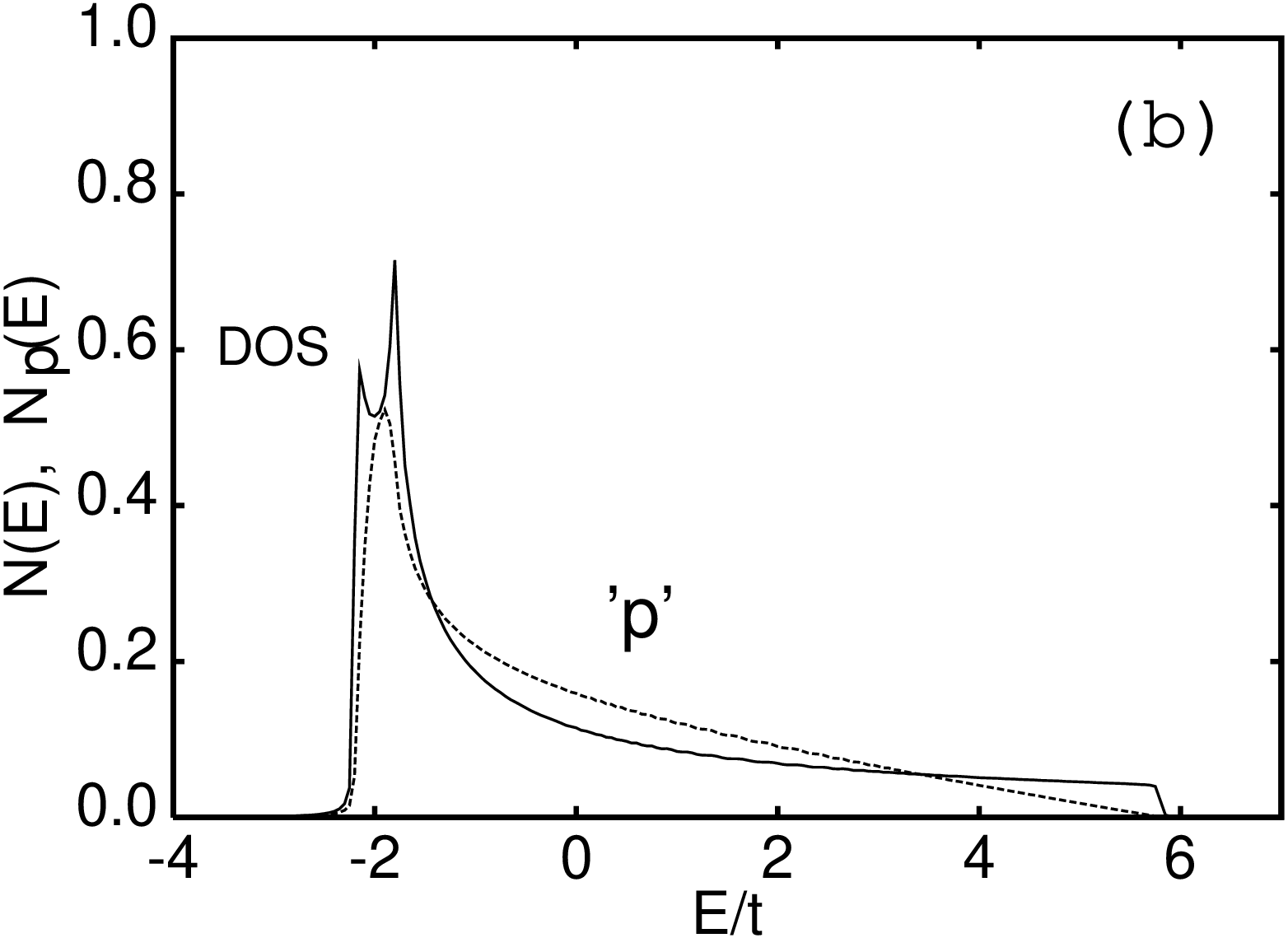}
\caption{ (a) The electron density of states $N(E)$
(full line)
and projected 
densities (dotted lines) for
extended $s$-wave $N_s(E)$ and  $d$-wave type $N_d(E)$ (a). $N(E)$
(full line) and 
$p$-wave type
$N_p(E)$ (dotted line)
for a normal
pure  system, the chemical potential $\mu=0$.
}
\end{figure}
\vspace{3cm}

\begin{figure}[htb]

\leavevmode

\vspace{-1.5cm}  
\epsfxsize=5.36cm
\hspace{2.65cm}
\epsffile{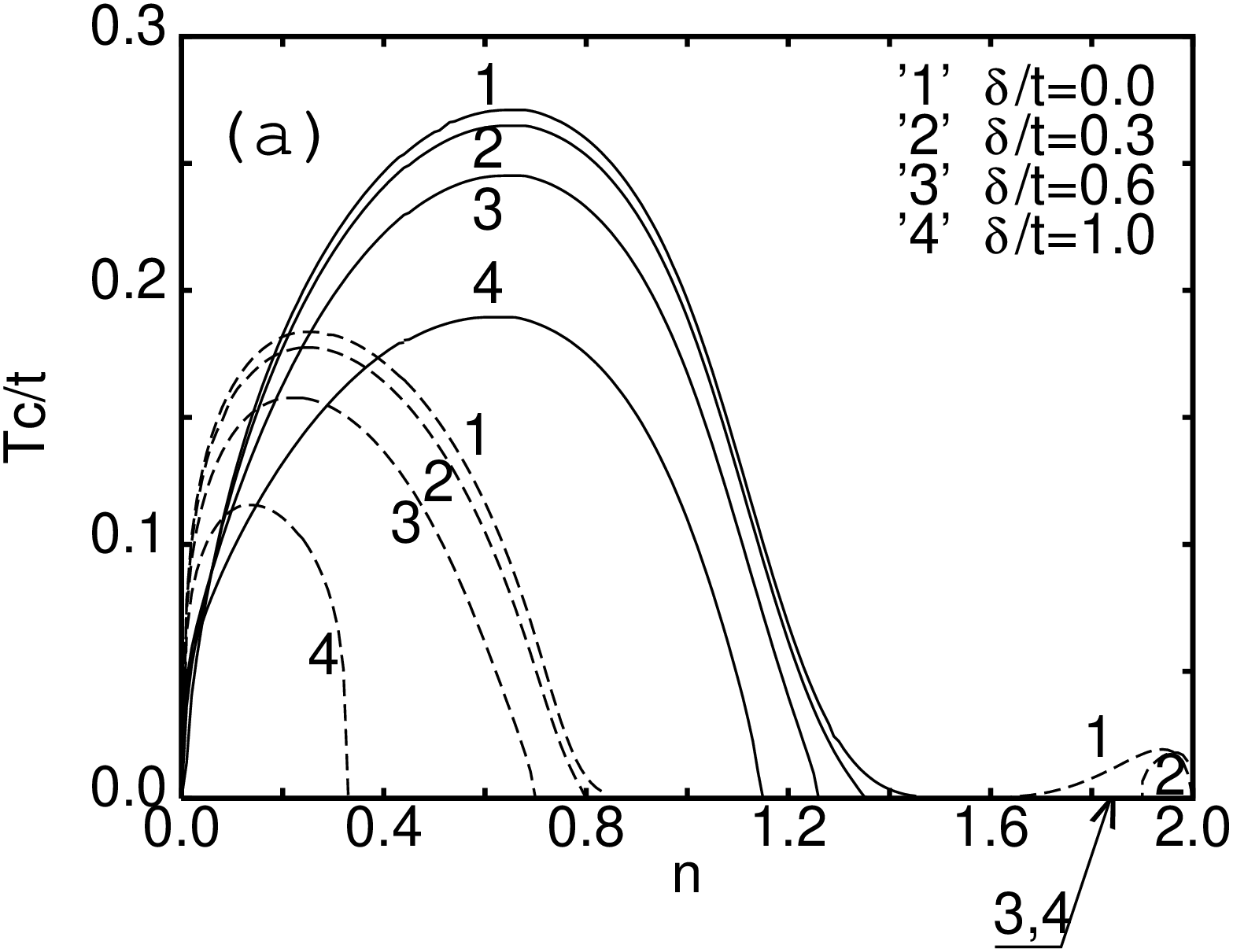}

\vspace{-1.8 cm}
\epsfxsize=5.5cm
\hspace{2.5cm}
\epsffile{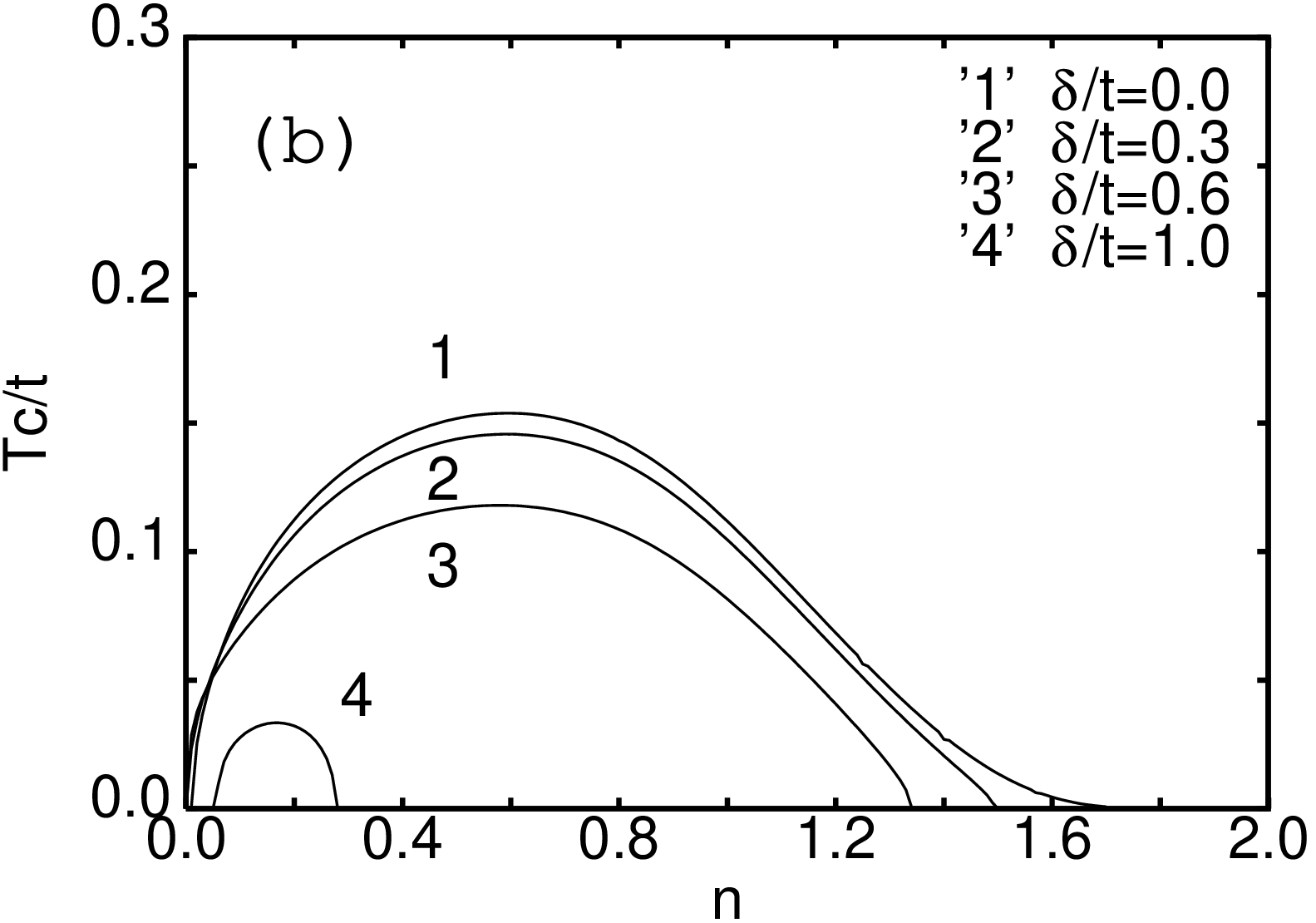}

\vspace{-1.8 cm}

\epsfxsize=5.36cm
\hspace{2.65cm}
\epsffile{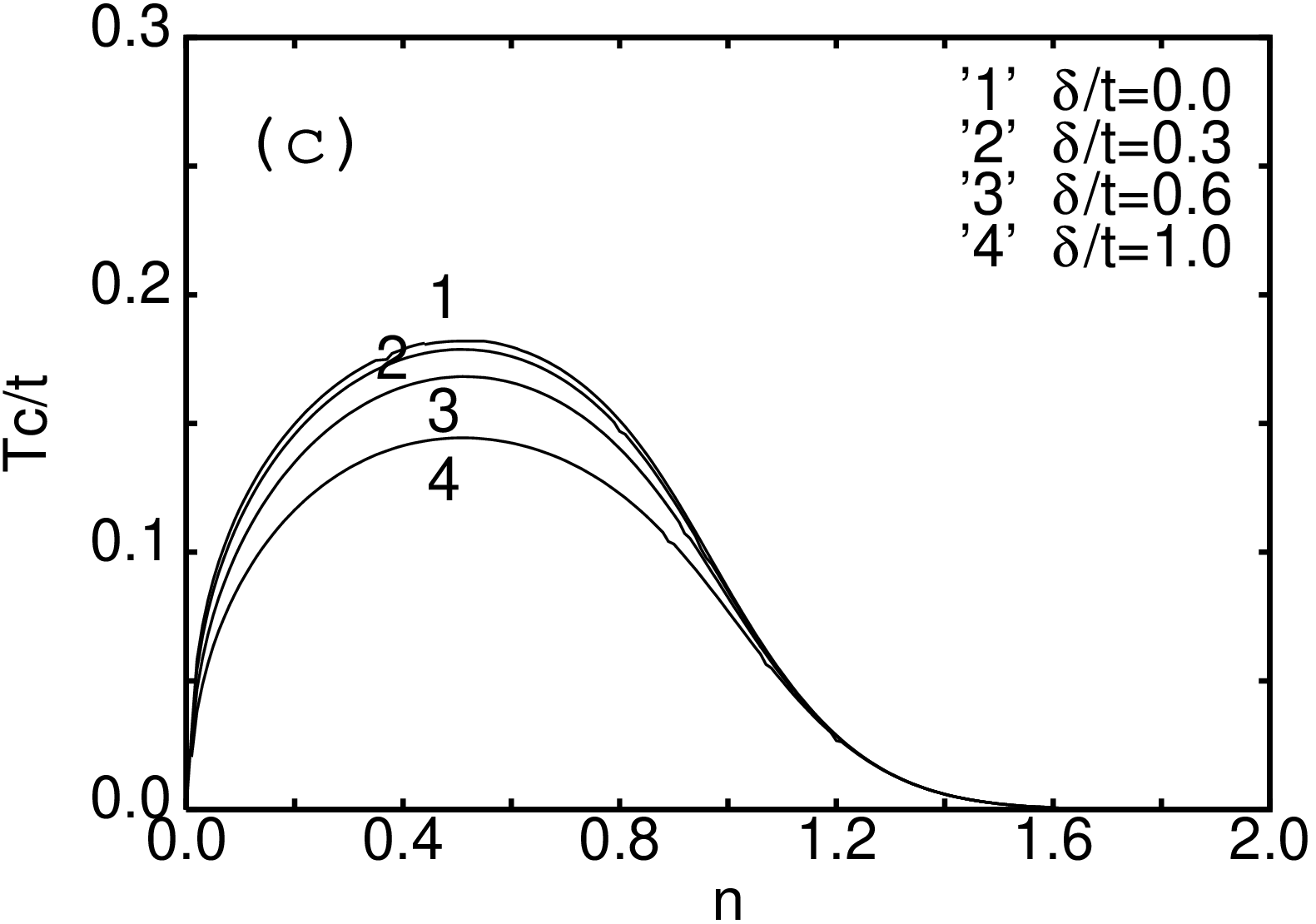}
\caption{(a) The
critical temperature $T_c$ versus band filling $n$ for extended s-wave and
 d-wave
pairing for a pure '1' and alloyed disordered system '2'--'4' (depending on
$\delta$); and $U=-1.5t$. Full lines
corresponds to $d$-wave solution while dashed extended $s$-wave ones.
(b) $T_c$ versus band filling $n$ for p-wave solution.
 (c) For
comparison $T_c(n)$ for on site
s-wave solution.}
\end{figure}   
\vspace{-1.2cm}

\begin{figure}[htb]
\leavevmode

\vspace{-2cm}

\epsfxsize=5.5cm
\hspace{-0.5cm}
\epsffile{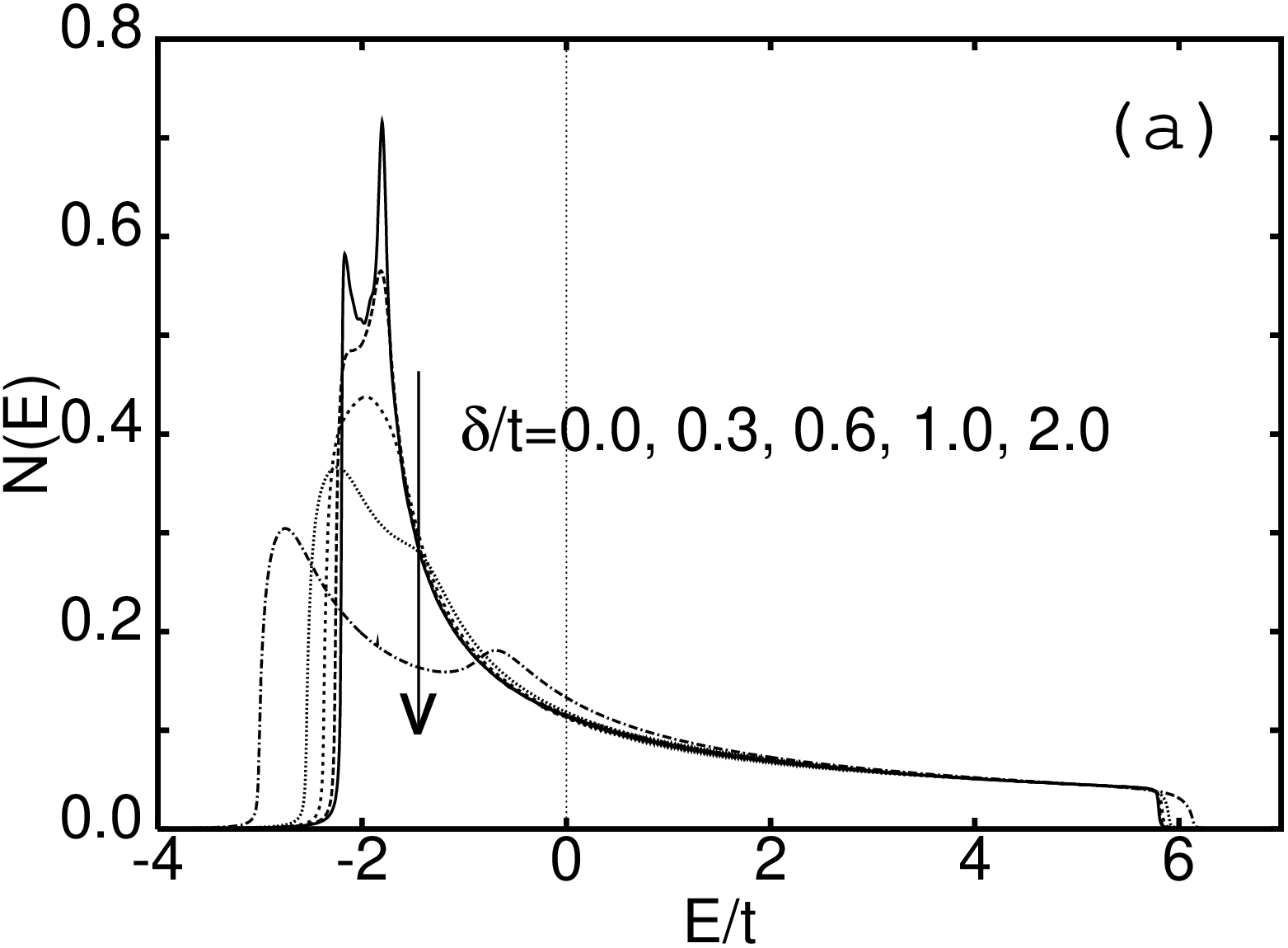}
\hspace{2.5cm}
\epsfxsize=5.5cm
\epsffile{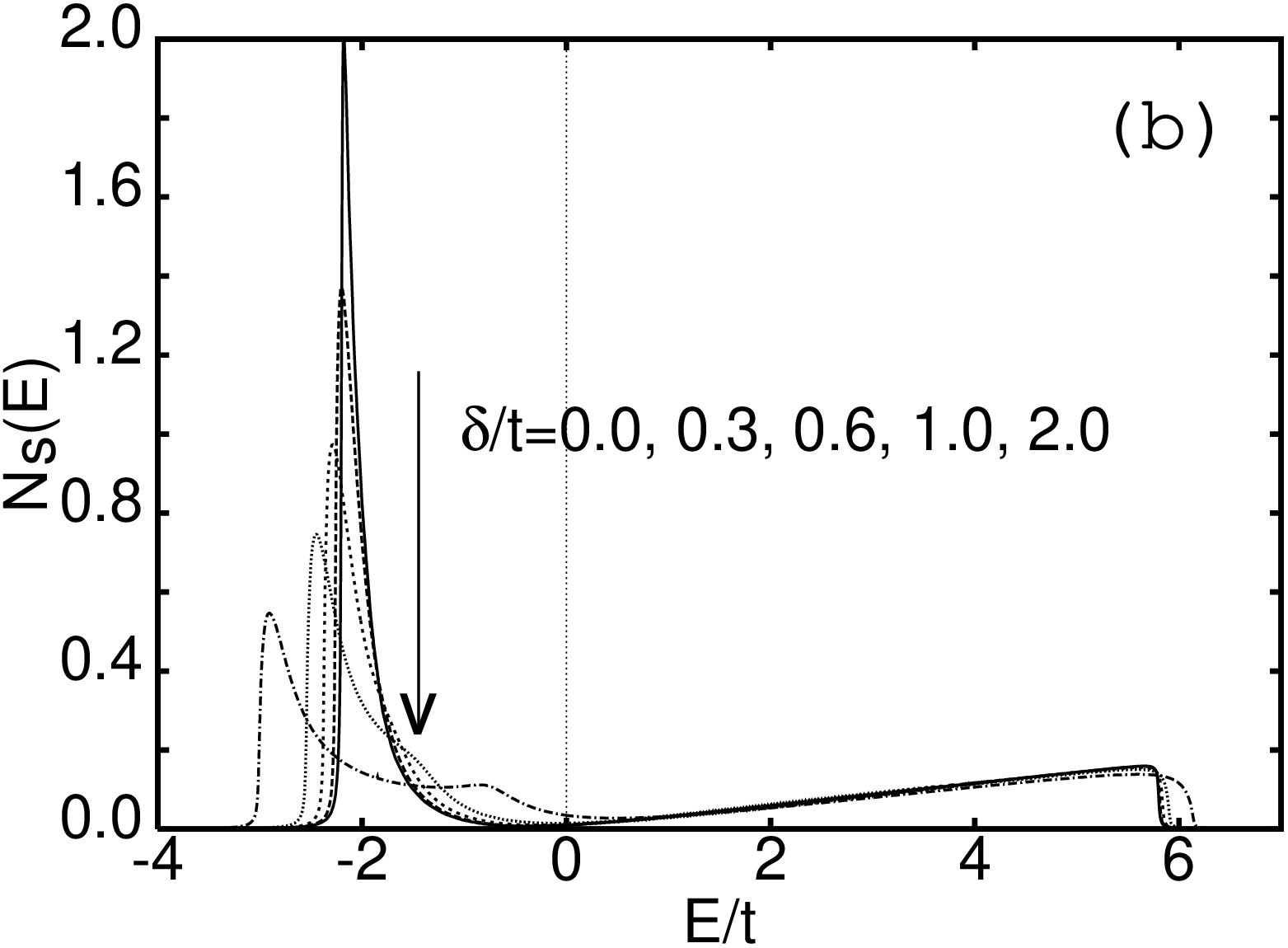}

\vspace*{-1.0cm}

\epsfxsize=5.5cm
\hspace{-0.5cm}
\epsffile{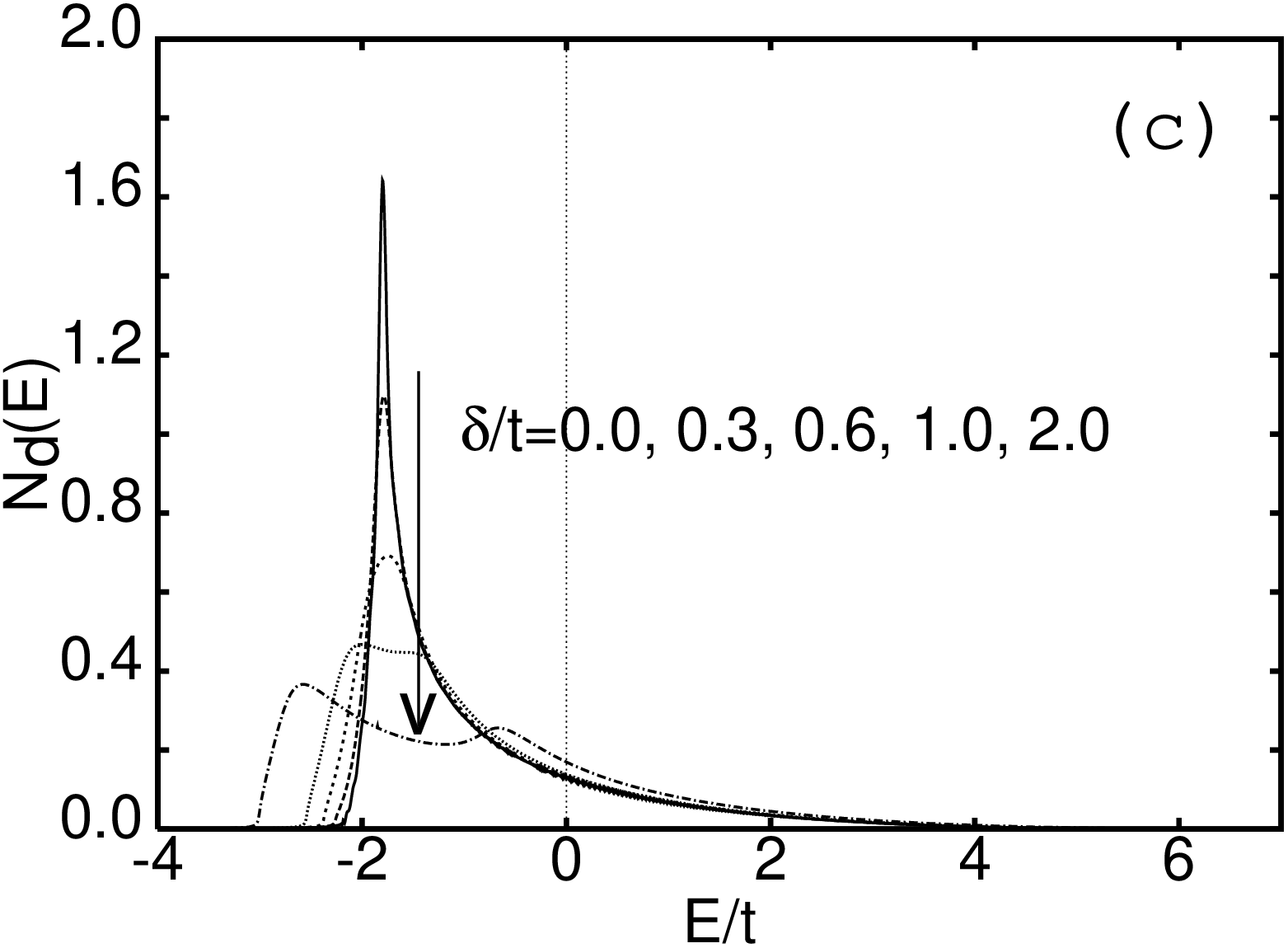}
\epsfxsize=5.5cm
\hspace{2.5cm}
\epsffile{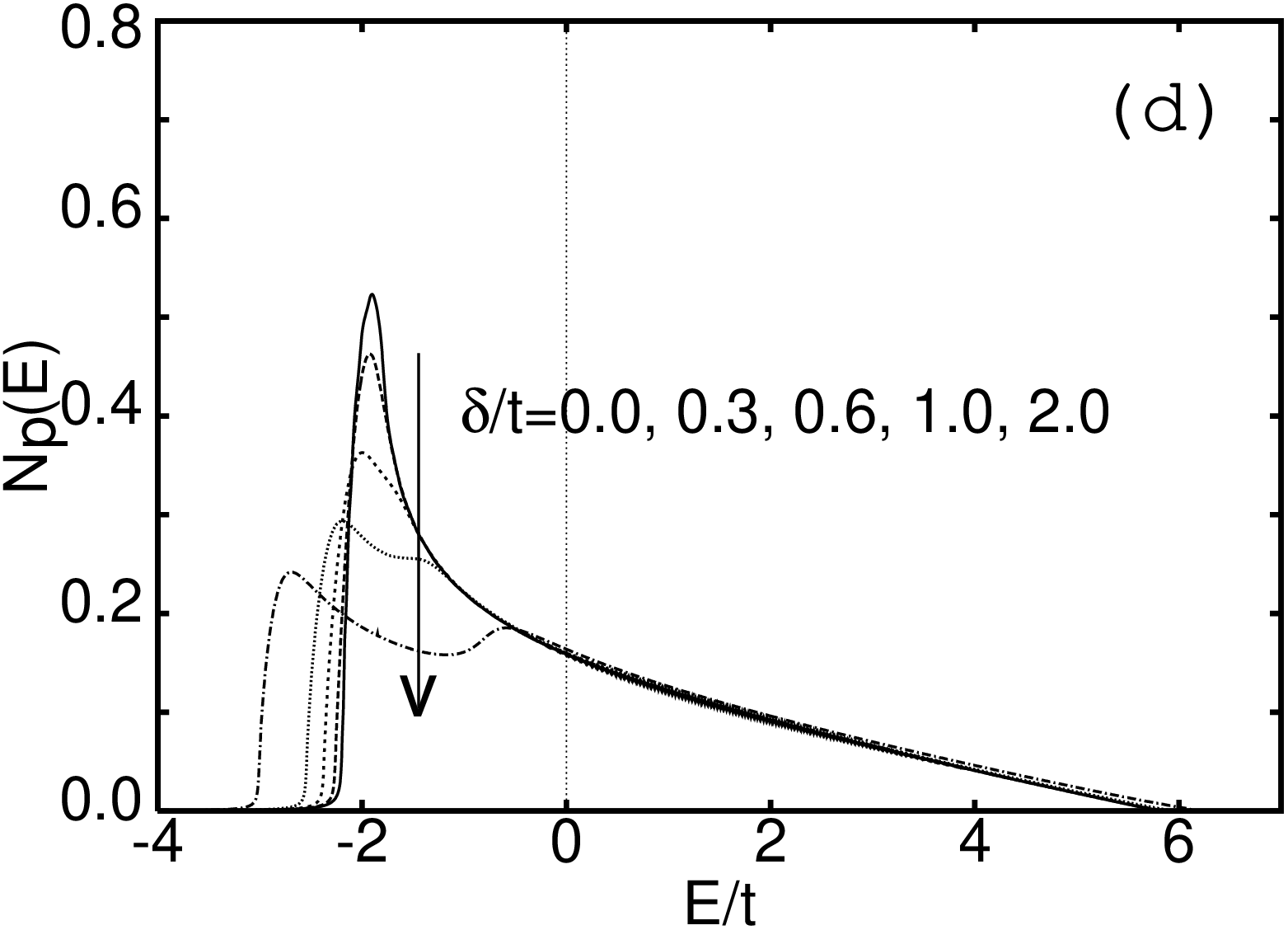}

\caption{ Normal state density of states $N(E)$ (a)
and projected   
densities:
extended s-wave $N_s(E)$ (b), d-wave type $N_d(E)$ (c) and p-wave
$N_p(E)$ (d),
respectively, for
different values of
disorder strength $\delta/t=0.0$, $0.1$, $0.2$, $0.3$. Arrows show
the directions of
$\delta$ change. Here, the chemical potential $\mu=0$.}
\end{figure}
\vspace{0cm}

\begin{figure}[htb]
\leavevmode
\epsfxsize=5.5cm
\hspace{2.5cm}
\epsffile{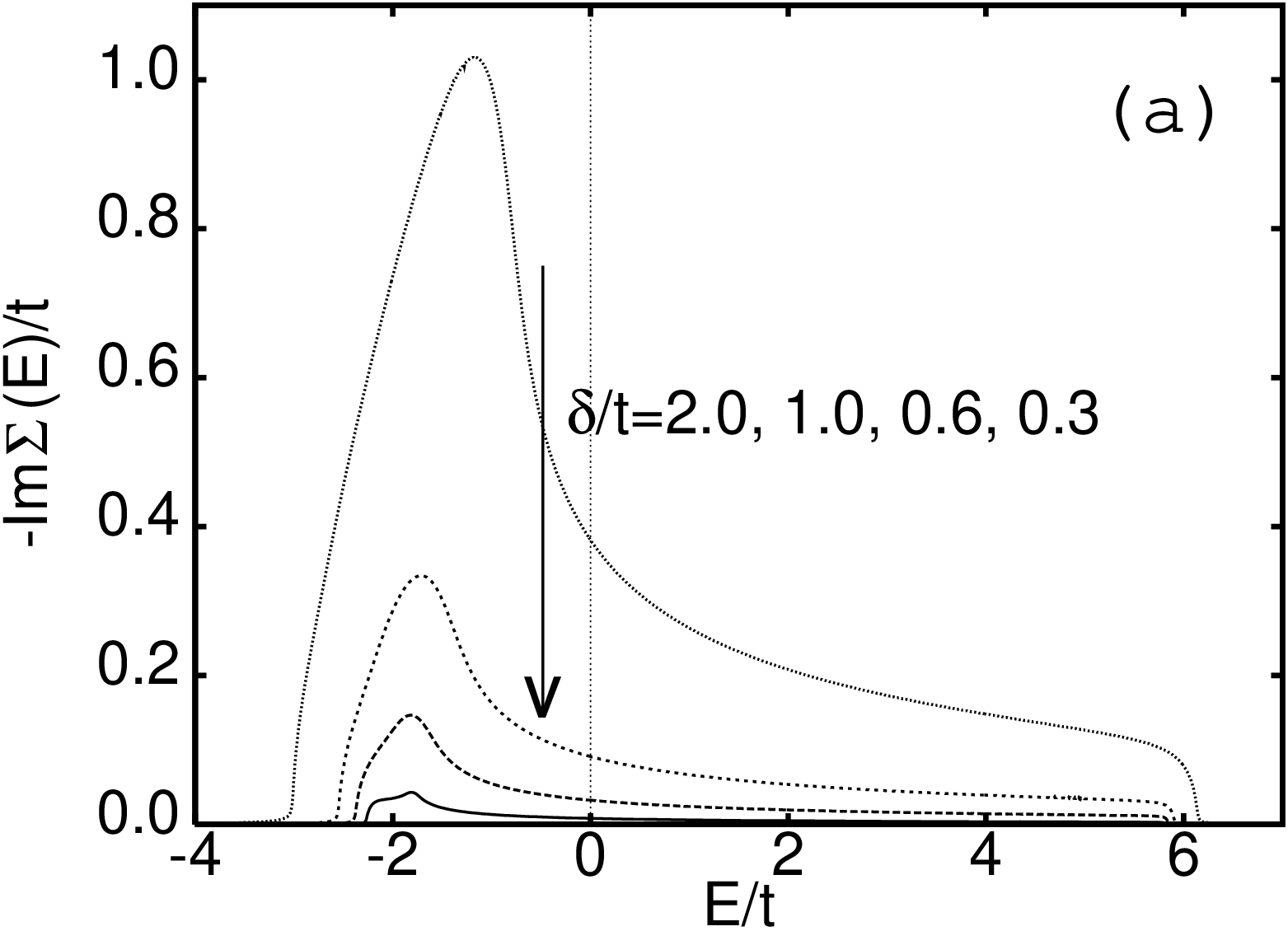}


\epsfxsize=5.5cm
\hspace{2.5cm}
\epsffile{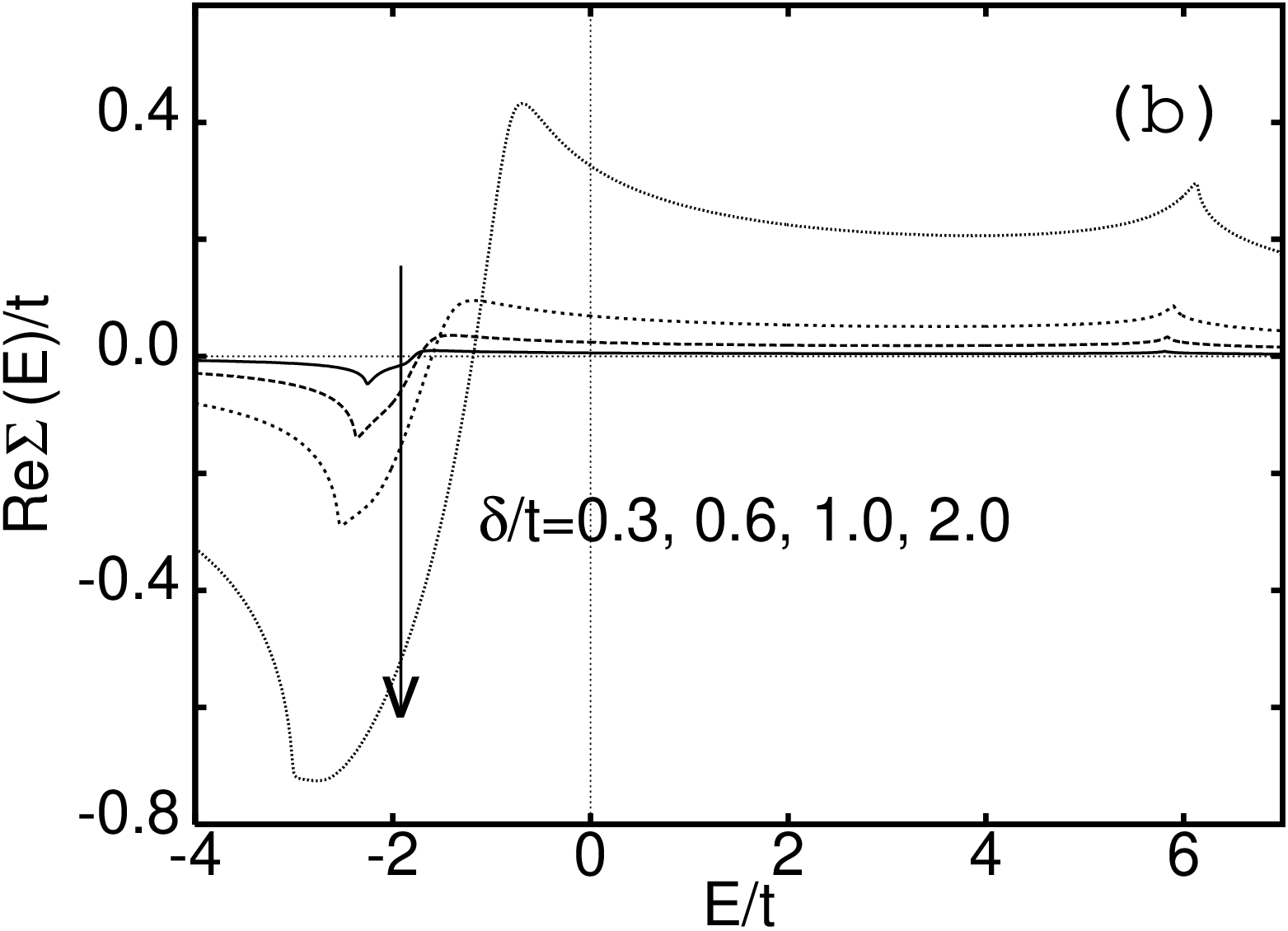}
\caption{ Normal state parts of  self
energy
$\Sigma(E)$: imaginary (a)
and real (b)  for different values of $\delta$.
}
\end{figure}
\vspace{0cm}

\begin{figure}[htb]
\leavevmode

\epsfxsize=5.5cm 
\hspace{2.5cm}
\epsffile{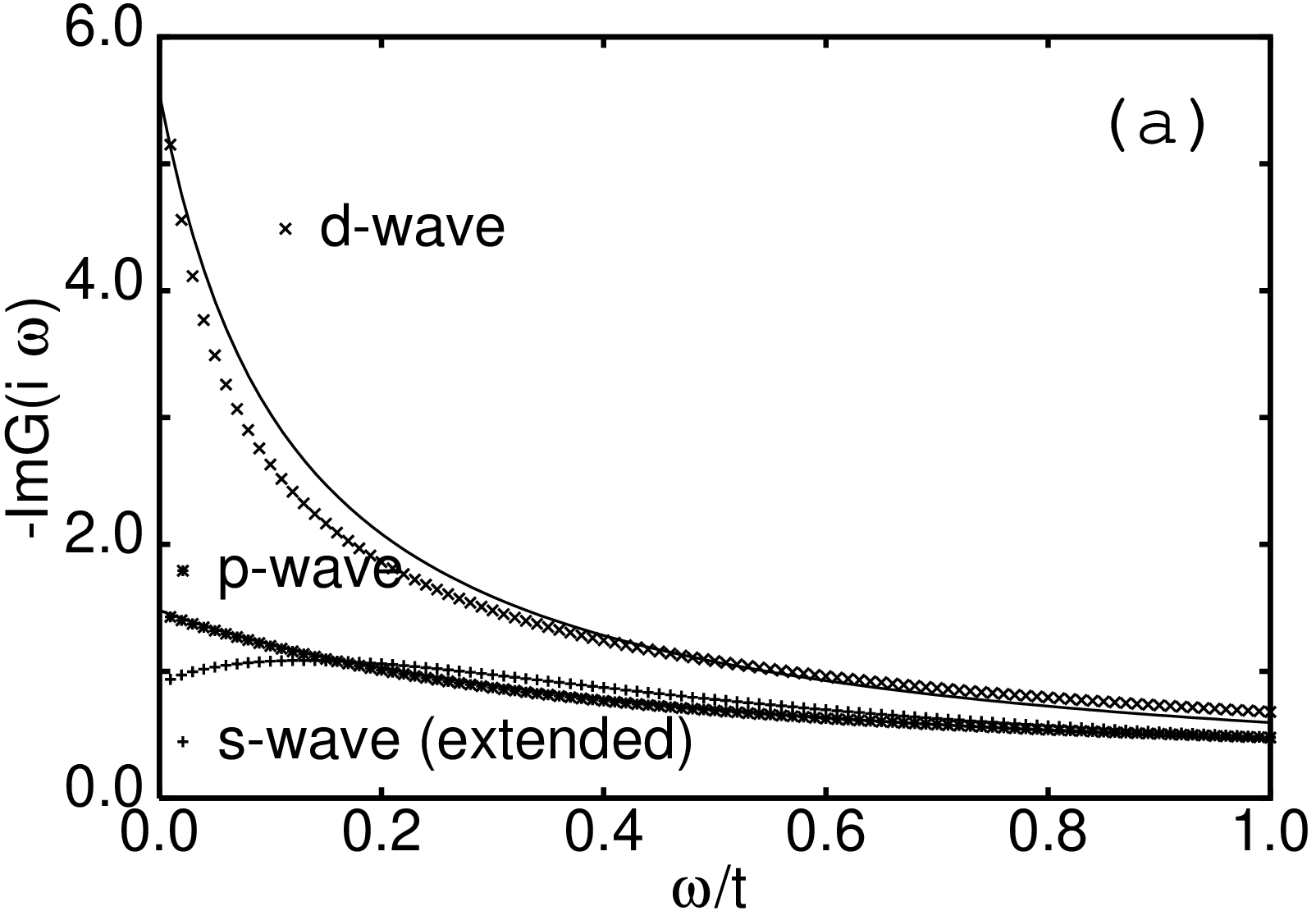}  

\vspace*{-1.5cm}

\epsfxsize=5.5cm
\hspace{2.5cm}
\epsffile{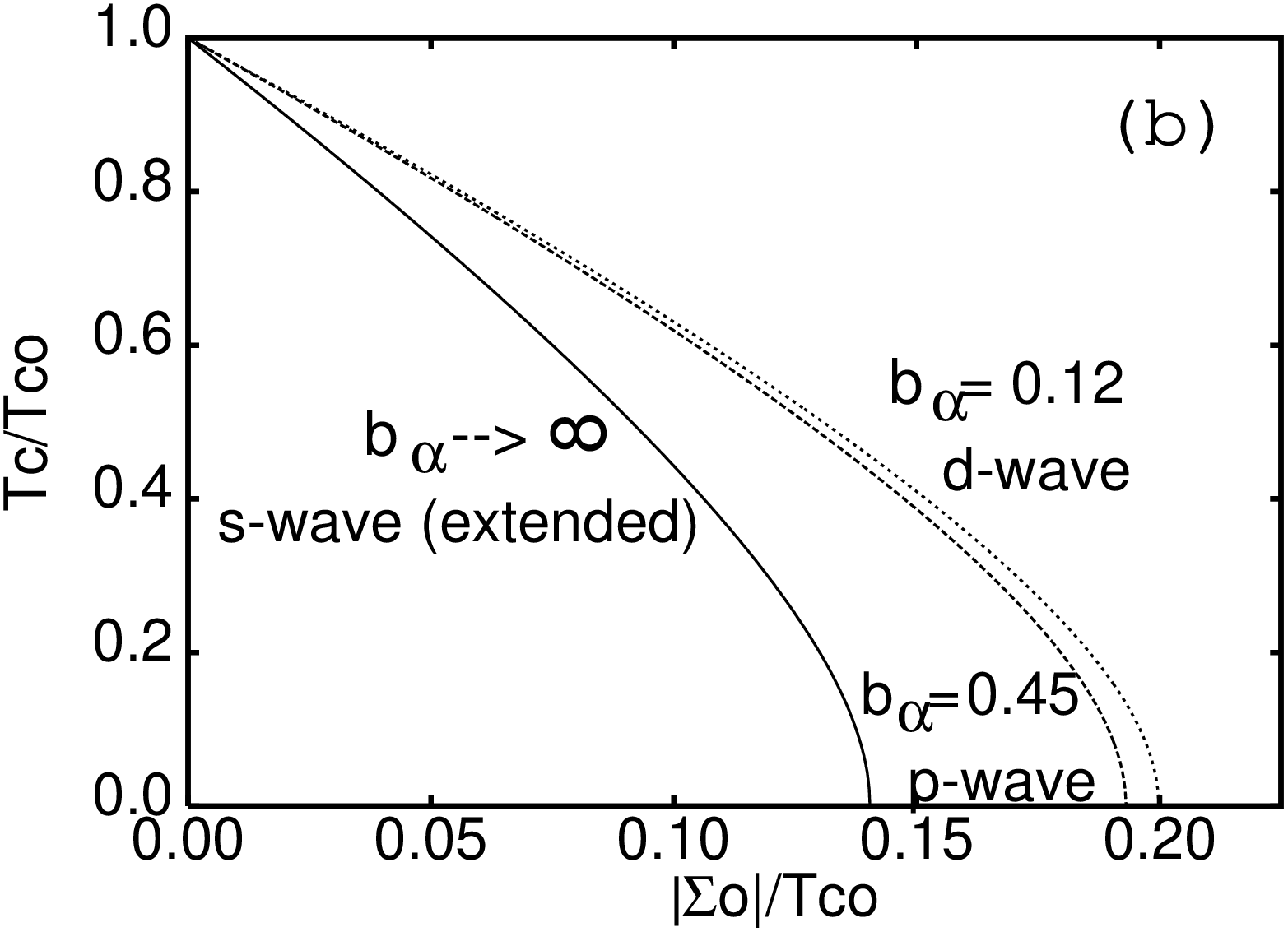}
\caption{ (a)
The imaginary part of Green functions: $-{\rm Im} G_d$, $-{\rm Im} G_p$ and
$-{\rm Im} G_s$
versus  imaginary energy $\imath \omega$. Fitted lines
$a_{\alpha}/(b_{\alpha}+\omega)$ are plotted for $d$--wave pairing where
$b_{\alpha}=0.12$ and  $p$--wave pairing $b_{\alpha}=0.12$
($a_{\alpha}=2/3$ for both curves).
(b) The critical temperature $T_c$
versus pair breaking parameter $|\Sigma_0|$ in the limit of weak disorder (both
$T_c$ and $|\Sigma_0|$ are
normalized to $T_{c0}$
of clean system) for few values of $b_{\alpha}$. The limit $b_{\alpha}
\rightarrow
\infty$  
corresponds to the standard Abrikosov--Gorkov formula (Eqs. 58--59).}
\end{figure}

\vspace{2cm}

\begin{figure}[htb]
\leavevmode
\epsfxsize=7.5cm
\hspace{3cm} \epsffile{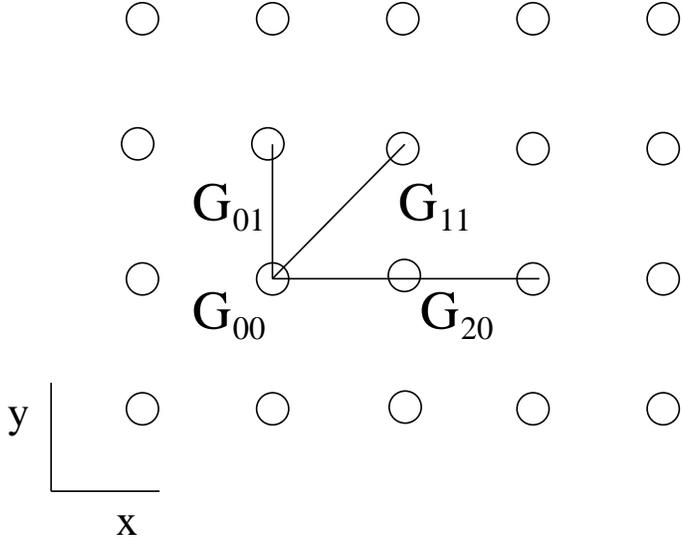}
\vspace{2cm}

\caption{Schematic picture of diagonal and off-diagonal Green functions
$G_{\alpha,\beta}=
G(\alpha \hat{\mb x} + \beta \hat{ \mb y} )$.}
\end{figure}

\end{document}